\documentclass{iucrjournals}
\usepackage{graphicx}
\usepackage{color,soul}
\title{Simulating Continuous-Rotation 3D Electron Diffraction: A Multislice and Bloch Wave Framework}

\author[a]{Małgorzata K. Cabaj%
	\IUCrAufn{These authors contributed equally to this work}%
	\IUCrEmaillink{mkcabaj@fzu.cz}%
	\IUCrOrcidlink{0000-0001-9184-4513}}
\author[c]{Jacob Madsen%
	\IUCrAufn[1]{}%
	\IUCrEmaillink{jacob.madsen@univie.ac}%
	\IUCrOrcidlink{0000-0001-5534-0553}}
\author[c]{Toma Susi%
	\IUCrEmaillink{toma.susi@univie.ac.at}%
	\IUCrOrcidlink{0000-0003-2513-573X}}
\author[a]{Lukáš Palatinus%
	\IUCrCemaillink{palat@fzu.cz}%
	\IUCrOrcidlink{0000-0002-8987-8164}}
\author[b]{Paul B. Klar%
	\IUCrCemaillink{paul.klar@uni-bremen.de}%
	\IUCrOrcidlink{0000-0002-1625-5353}}

\affil[a]{Institute of Physics, Czech Academy of Sciences, Prague, Czechia}
\affil[b]{Faculty of Geosciences, University of Bremen, Bremen, Germany}
\affil[c]{Faculty of Physics, University of Vienna, Vienna, Austria}

\begin{document} 
\maketitle 

\begin{synopsis}
This work establishes an approach for simulating continuous-rotation 3D electron diffraction (3D ED) data employing either Bloch wave or multislice calculations. The study investigates relevant parameters of the modelled experiment. Results are cross validated by comparing multislice with Bloch Wave calculations. The approach provides a clear strategy for generating data sets suitable for crystallographic analysis.
\end{synopsis}

\begin{abstract}
To improve the agreement between measured and calculated intensities of three-dimensional electron diffraction (3D ED) experiments, a simulation pipeline is needed to assess and quantify the influence of various structural and experimental parameters. We present a computational pipeline, built upon the \emph{ab}TEM Python package, to simulate continuous-rotation 3D electron diffraction data based on either the Bloch wave or the multislice formalism. Multislice calculations in arbitrary orientations are achieved through large supercells and windowing. We investigate the convergence of key simulation parameters and test their consistency, establishing suitable parameters for accurate and efficient simulations. The pipeline's applicability and robustness are demonstrated through case studies on cubic silicon as well as stretched and sheared variants, analyzing the influence of electron kinetic energy, sample thickness, orientation, and symmetry on simulated diffraction intensities. Finally, we investigate a representative selection of seven compounds, including cubic $\mathrm{SrTiO_3}$, monoclinic $\alpha$-glycine $\mathrm{C_2H_5NO_2}$, and triclinic kyanite $\mathrm{Al_2SiO_5}$ to validate the method. This framework for the simulation of 3D electron diffraction data establishes an approach to investigate the dependence of diffracted intensities on experimentally relevant parameters which are difficult to systematically investigate in experiments.
\end{abstract}


\section{Introduction}

Electron diffraction probes the atomic structure of crystalline materials. The diffraction patterns emerge from the interaction of high-energy electrons with the periodic potential of the material. An important development in this field is three-dimensional electron diffraction (3D ED), a method based on a series of diffraction patterns measured at a usually continuous range of crystal orientations. The analysis of 3D ED data sets provides comprehensive insights into the symmetry, unit cell parameters, average structure and, to some extent, the morphology of the crystal.  The main application of 3D ED is the solution or refinement of crystal structures from crystallites that are too small for typical single-crystal X-ray diffraction (XRD) experiments \cite{Mugnaioli2009,Gemmi2019}.

Despite the continuous improvements of 3D ED methodology, residual factors (\emph{R}-factors) obtained during refinement generally remain higher than those typical of XRD, representing a prominent shortcoming in electron crystallography. Although a few exceptions report \emph{R}-factors as low as expected for XRD \cite{Suresh2024, Oyonarte2025}, most authors only offer tentative, qualitative explanations for high \emph{R}-factors. High \emph{R}-factors are frequently attributed to unmodeled physical aspects, such as multiple scattering, inelastic scattering, crystal imperfections, or overall data quality \cite{Broadhurst2020, Klar2023, Mendis2024a, Mendis2024b, Poppe2024}. However, a quantitative assessment of how these systematic errors impact structure analysis is rarely provided. This is likely because a simple relationship to estimate these effects is not expected, and a suitable, overarching methodology is currently lacking. Ultimately, a deeper understanding of the factors influencing diffracted intensities is essential to reduce systematic errors, lower \emph{R}-factors, and yield more precise structure models with clearer Fourier maps.

While several studies have explored how the choice of form factors \cite{Gruza2019, Suresh2024}, dynamical diffraction \cite{Palatinus2015a, Palatinus2015b, Klar2023}, or diffraction geometry \cite{Saha2022, Klar2023, Plana-Ruiz2025, Schmitt2026} affect 3D ED refinements, the literature lacks a systematic analysis across a broad range of materials. More importantly, there is a critical absence of studies that directly isolate certain experimental parameters to determine their singular impacts on final structure models. For instance, while it is known that inelastic scattering such as phonon and plasmon excitations contribute to the diffuse background and alter Bragg reflection intensities \cite{Mendis2024a, Mendis2024b}, their precise structural consequences remain poorly understood. Similarly, the effects of crystal morphology and internal imperfections (e.g., point defects, dislocations, strain, mosaicity, and small-angle grain boundaries) on diffracted intensities have yet to be systematically investigated.

To address these limitations, this work aims to establish a general approach that uses realistic simulation of diffraction data to evaluate and quantify these complicated effects. By doing so, we intend to provide a foundation that, in the long term, mitigates this lack of understanding and allows for the accurate estimation and reduction of systematic errors in 3D ED data analysis.

The simulation of 3D ED data allows exploration of a wide range of structural configurations without the constraints of material availability or experimental setup limitations. Furthermore, simulations provide precise control over parameters such as electron wavelength, crystal morphology, sample orientation, and crystal imperfections. In experiments, full control of all these parameters is challenging or even impossible.

The two primary formalisms used to calculate electron diffraction patterns are the Bloch wave \cite{Bethe1928} and the multislice \cite{Cowley1995} approaches, both based on solving the Schrödinger equation with relativistically corrected electron mass and wavelength. The Bloch wave formalism uses the periodic potential of the crystal and treats the electron wavefunction as a superposition of Bloch waves. This method excels in calculating diffracted intensities for perfect crystals in arbitrary orientations. The multislice method approximates the propagation of an electron wave through the crystal by dividing the electrostatic potential into a series of thin slices and applying a sequence of phase shifts and Fresnel propagation steps. This approach is typically used for zone-axis oriented samples and is well-suited for handling complex structures, including those with defects, interfaces, and amorphous regions, as it allows for more flexible modeling of non-periodic potentials and inelastic scattering processes.

Soon after multislice was established, it was shown that, under certain conditions, calculated intensities are expected to be identical to those calculated with the Bloch wave approach \cite{Goodman1974, Ma1990, Bangun2024}. There are a few comparative studies exploring the sources of differences of both method outputs, how these differences influence the outcomes and which formalism will be better suited to answer a particular question \cite{Koch2000, Lobato2015, Lubk2015, Yang2017}.

To our best knowledge, until now, all the studies investigating or comparing the accuracy of Bloch wave and multislice methods were limited to zone-axis patterns or tilt angles of a few miliradians. This preference can be understood from an imaging point of view as images obtained with a transmission electron microscope (TEM) or scanning TEM (STEM) are much easier to interpret when columns of atoms are aligned with the electron beam. Despite advances in computational methods and the increase in computational power, the suitability of the algorithms and associated tools for the simulation of a full 3D ED data series has never been investigated.

Numerous software packages have been developed and are actively extended with the goal to simulate high-resolution TEM images, STEM images, and diffraction patterns. Among the available programs and packages there is, e.g., \emph{ab}TEM \cite{abTEM}, py4DSTEM \cite{py4DSTEM}, MuSTEM \cite{MuSTEM}, MULTEM \cite{MULTEM}, Prismatic \cite{Prismatic}, Dr. Probe \cite{drProbe} or ToTEM \cite{ToTEM}. Most packages implement either the Bloch wave approach or the multislice approach, and only few packages have both approaches implemented, namely py4DSTEM, JEMS \cite{JEMS}, eMAP with eSLICE \cite{eMAP} although the latter packages require commercial license and are not open source. Even with these programs, it seems that a rigorous comparison has not been reported, possibly because the implementations make it difficult to use them together in a consistent manner. Furthermore, multislice has until now been considered unable to handle arbitrary sample rotations.

This work presents an approach to simulating 3D ED data sets using either the Bloch wave or multislice approaches based on a common set of input parameters. The effects of critical simulation parameters on calculated intensities are investigated with the goal to validate the approach. The convergence of these parameters is also assessed by comparing results obtained with the Bloch wave approach with results obtained with the multislice approach. Data sets are generated for a range of compounds to assess the suitability and reliability of the approach for low to high-symmetry and low to high-Z compounds. Later, the implications of this work and practical aspects are discussed.

\section{Methods}

Our simulation setup uses a custom pipeline built in Python that makes use of the open-source \emph{ab}TEM package \cite{abTEM}. This setup was designed to simulate selected functions of an electron microscope or electron diffractometer, so that diffraction patterns from any crystal orientation can be calculated. The simulations were carried out using two main methods: the Bloch wave formalism and the multislice approach. While the simulations attempt to be as realistic as possible, several simplifications are still applied, which may be removed in the future. First, we do not include inelastic scattering due to, e.g., phonon or plasmon excitations, which contribute to the diffuse background and affect Bragg intensities in experimental data. Hence, the simulations only consider the elastic part of the electron's interaction with the sample. A perfect crystal made of independent atoms with a thickness independent of the crystal orientation is assumed. Thermal displacements are modelled with a space- and time-averaged Debye-Waller factor. Thus, thermal diffuse scattering is not modelled and effects of frozen phonons are currently omitted.

Section~\ref{sim_par} presents a detailed analysis of the convergence of simulation parameters and Section~\ref{sim_Si} includes a systematic comparison to confirm that meaningful results are obtained with both methods.

\subsection{Computational Pipeline}

\begin{figure}[!h] %
	\begin{center}
		\includegraphics[width=0.5\textwidth]{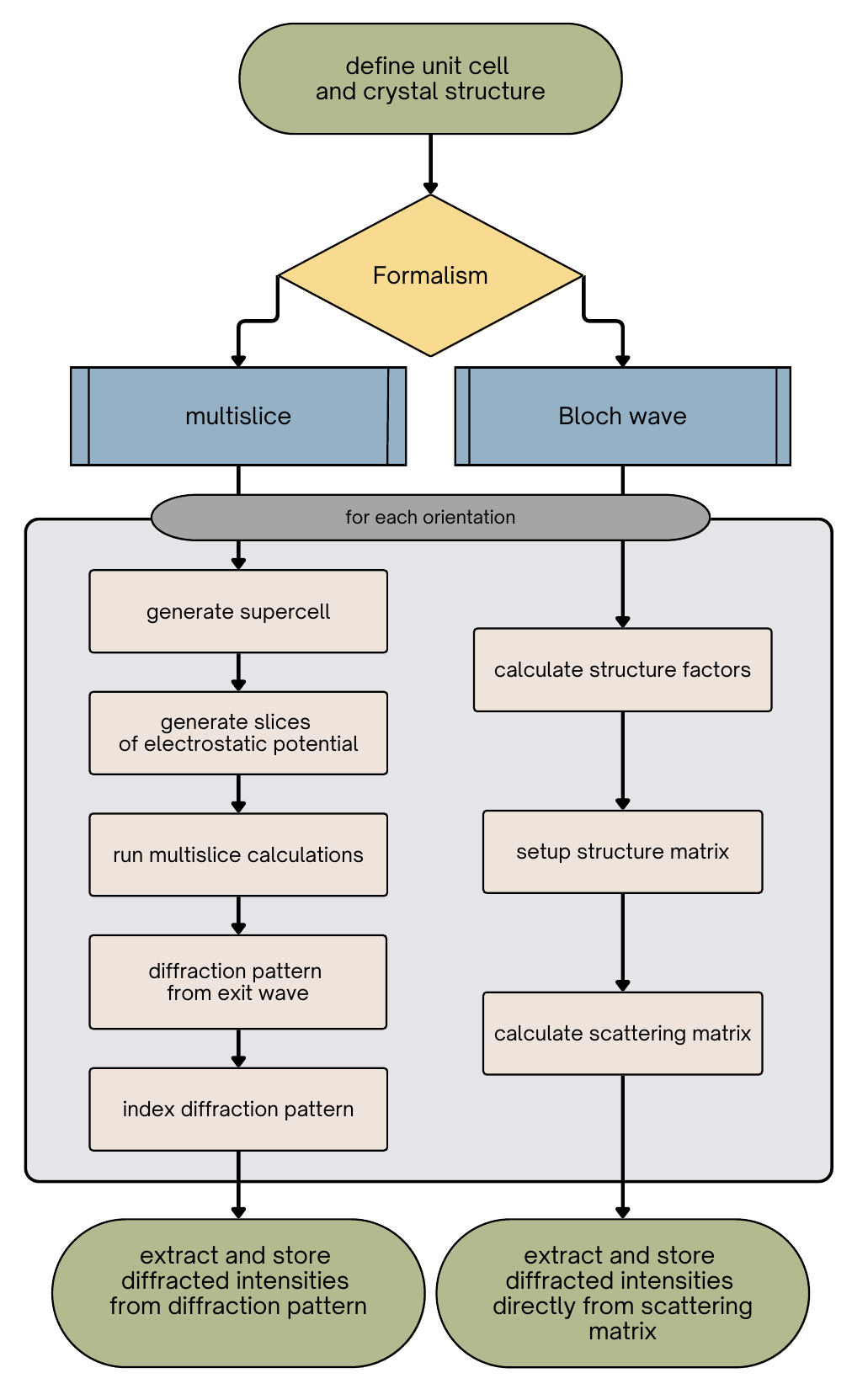}
	\end{center}
	\caption{Steps needed to obtain simulated intensities with both formalisms.} 
	\label{fig:Flowchart}
\end{figure}

The pipeline is implemented as a Python package called \emph{py3DED} and consists of a set of tools that interact with the \emph{ab}TEM package. In a nutshell, the simulation of a 3D ED data set consists of defining a crystal structure, simulating the propagation of the electron beam through the crystal for a range of orientations, and storing the diffracted intensities for each orientation as a function of propagation depth. The flow chart in Fig.~\ref{fig:Flowchart} illustrates the most relevant steps of the simulation.

Crystal structure models are defined by reading a crystallographic information file (CIF)  using the \emph{Atomic Simulation Environment} package (ASE) \cite{Larsen2017}. Structure factors and the calculation of the electrostatic potential from an arrangement of atoms use the form factor parameterisation defined by Lobato \& van Dyck \cite{Lobato2014}. These form factors are multiplied by an isotropic Debye-Waller factor $\exp [ -8 \pi^2 \langle u^2 \rangle \frac{\sin^2(\theta)}{\lambda^2} ] = \exp[ -2 \pi^2 \langle u^2 \rangle \left|\mathbf{g} \right|^2 ]$ where $\langle u^2 \rangle = U_\mathrm{iso}$ is the isotropic mean-squared displacement as typically used in crystallography and within the harmonic approximation describes the variance $ \sigma^2 $ of a Gaussian displacement distribution. Throughout this study, a displacement parameter $ U_\mathrm{iso} = \sigma^2 = 0.01~\mbox{\AA}^2 $ was imposed on all atoms for Bloch wave as well as multislice calculations.

For the Bloch wave approach, the steps are analogous to most standard implementations of the method, e.g. in the Bloch wave program Dyngo \cite{Palatinus2015b}. With this method, the exit wave $\mathbf{\varphi}(t)$ of a crystal with a thickness $t$ is computed via the scattering matrix $\mathbf{S}$ according to $\mathbf{\varphi}(t) = \mathbf{S} \mathbf{\varphi}(0)$, where $\mathbf{\varphi}(0)$ represents the incident beam amplitudes. Under plane-wave illumination conditions, the incident amplitude $\mathbf{\varphi}_h(0) = 1$ for the primary beam ($hkl = 000$) and is zero for all other beams $\mathbf{h}$.
The scattering matrix $\mathbf{S}$ accounts for dynamical electron scattering within the crystal and is formulated as:

$$\mathbf{S} = \mathbf{M} \exp \left( \frac{2\pi i t}{2K} \mathbf{A} \right) \mathbf{M}^{-1}$$

Here, $t$ is the crystal thickness and $K$ is the magnitude of the incident wavevector. The matrices $\mathbf{A}$ and $\mathbf{M}$ are defined as follows: $\mathbf{A}$ is the structure matrix, which couples the different diffracted beams. Its elements are defined by the crystal potential and the deviation from the exact Bragg condition. The crucial component of the diagonal elements $A_\mathbf{gg} = \frac{2K S_g}{\sqrt{1 + g_z/K}}$ are the excitation errors $S_g$. $g_z$ is the $z$-component of the reciprocal lattice vector $\mathbf{g}$. Off-diagonal elements $A_\mathbf{gh} = \frac{U_\mathbf{g-h}}{\sqrt{1 + g_z/K}\sqrt{1 + h_z/K}}$ describe the dynamical coupling between beams $\mathbf{g}$ and $\mathbf{h}$. 

% Lukas comment, 21 July 2026
The selection of reflections used to build the structure matrix $\mathbf{A}$ is controlled by two parameters, the resolution limit $g_{\mathrm{max}}$ and the excitation error limit $S_{g_{\mathrm{max}}}$. The excitation error $S_{g}$ can be interpreted as the distance of a reflection \textbf{h} from the Ewald sphere in units of $\mbox{\AA}^{-1}$. $S_{g}$ is positive for reflections inside the Ewald sphere, $0~\mbox{\AA}^{-1}$ if the Bragg condition is exactly fulfilled, and negative for reflections outside the Ewald sphere. The resolution limit $g_{\mathrm{max}}$ defines the maximum length of the reciprocal lattice vector $\mathbf{g}$. Reflections that satisfy both criteria, i.e., $g_\mathbf{h} \leq g_{\mathrm{max}}$ and $\left|S_{g} \right| \leq S_{g_{\mathrm{max}}}$, are included in the structure matrix \textbf{A}, while those that do not are excluded.

$U_\mathbf{g-h}$ is the Fourier coefficient of the electrostatic potential, which is proportional to the structure factor $\mathbf{g}-\mathbf{h}$ based on the average structure. The relationship between $F_{\mathbf{h}}$ and $U_{\mathbf{h}}$ is described by $F_{\mathbf{h}} = \frac{h^2}{2 m |e|} \cdot U_{\mathbf{h}} \cdot V_{UC}$, where $h$ is the Planck constant, $m$ is the relativistic electron mass, $e$ is the unit charge, and $V_{UC}$ is the unit cell volume. $\mathbf{M}$ is a diagonal matrix with diagonal elements $m_{ii} = \frac{1}{\sqrt{1 + g_{z,i}/K}}$. The scattering matrix $\mathbf{S}$ is calculated from the structure matrix and directly yields the diffracted intensities $I_\mathbf{h}$. From the shown formulas it is clear that the relationship between structure factors $F_\mathbf{h}$ and diffracted intensities $I_\mathbf{h}$ is highly non-linear so that in general $I_\mathbf{h} \sim |U_\mathbf{h}|^2$ does not hold.

 By adapting the selection of reflections to a given orientation, the Bloch wave procedure is iterated over a list of predefined orientations. For each orientation, intensities are calculated for a defined range of propagation depths so that different crystal thicknesses are simulated. As the computational cost increases with the number of reflections contributing to the structure matrix, zone axis patterns tend to be more expensive than arbitrary orientations with less reflections, especially if $S_{g_{\mathrm{max}}}$ is small.

The equivalent setup for the multislice case is substantially different. Here, the crystal structure is rotated to a desired orientation, and fills a simulation box with fixed dimensions $L_{x} \times L_{y} \times L_{z}$. For all atoms inside the box, the electrostatic potential is calculated and divided into slabs stacked parallel to the propagation direction $z$ of the incoming plane wave. The potential of each slab is then projected onto a 2D slice with dimensions $L_{x} \times L_{y}$. The incoming plane wave propagates through the first slice and is modulated by the 2D electrostatic potential. The state of the wave after the first slice is described by the convolution of the Fresnel free-space propagator with the transmission function multiplied by the initial wave function. The resulting wave function is then used for the next iteration, which is repeated until the end of the simulation box is reached and the wave has propagated through a total crystal thickness of $L_{z}$. This approach, which is based on fast Fourier transforms for computational efficiency, implicitly imposes periodic boundary conditions on the simulation box.

For the determination of diffraction pattern intensities at selected intermediate steps and after the final propagation step, the wave function $\psi(x,y)$ is Fourier-transformed and squared to yield the diffraction pattern corresponding to a certain crystal thickness and orientation. However, as the box size is kept fixed and not adapted to the structure or orientation, atoms do not align with the periodic boundary conditions, which is also known as wrap-around error. To minimise its effect on the Fourier transform and thus on the diffraction pattern, a two-dimensional Hann window $w(x,y) = \sin^2(\pi\frac{x}{L_x}) \cdot \sin^2(\pi\frac{y}{L_y})$ was applied to the two-dimensional wave function $\psi(x,y)$ \cite{Blackman1958}. Note that the windowing function is only applied to the wave function before the Fourier transform and does not affect the propagation of the wave function through the crystal. The windowing function reduces the amplitudes of the Fourier transform by a constant factor, but does not change their relative intensities (see SI Section \ref{Windowing}).

Diffracted intensities are not necessarily restricted to a single grid point or pixel. Therefore, diffracted intensities $I_{\mathbf{h}}$ are extracted by first indexing the diffraction pattern and then summing up all pixels within a radius of $0.02~\mbox{\AA}^{-1}$ in the proximity of the nominal reflection position. Due to the requirement of the large simulation box, multislice calculations are significantly more expensive than Bloch wave calculations. Even though the diffraction patterns are calculated before the indexation and intensity extraction step, here diffraction patterns are not stored as this would occupy a lot of disk space. 

For both Bloch wave and multislice calculations, a simulated data set consists of diffracted intensities $I_{\mathbf{h}}$ as a function of discrete orientation angle $\alpha$ and propagation depth $z$. Reflection intensity data are stored alongside relevant metadata as a three-dimensional ($\mathbf{h}$, $\alpha$, $z$) array, which results in a typical data set size of a few GB. 

For a given thickness and reflection, the data subset $I_{\mathbf{h}}(\alpha)$ describes the profile, also known as rocking curve, of a reflection as a function of the rotation parameter $\alpha$. Its integration yields integrated intensities, which correspond to the quantities that are typically used for structure model refinement within the dynamical theory of diffraction \cite{Palatinus2019, Petříček2023}. For structure analysis within the kinematical theory of diffraction, a geometry correction (Lorentz correction) needs to be applied \cite{Zhang2010} so that diffracted intensities are proportional to structure factor amplitudes.

For orthogonal unit cells, an alternative approach exists for the simulation of data sets with multislice calculations that fulfills periodic boundary conditions and thus leverages the requirements for large simulation boxes. For all zone axes $[uvw]$ with integer $u$, $v$, and $w$, the propagation direction is parallel to a lattice vector and a periodic cell can be used as the simulation box. Small orientational deviations from a zone axis can then be modelled by applying a linear phase ramp to the wave function at each slice, effectively shifting the origin of each slice and thereby allowing the simulation of tilted orientations \cite{Chen1997}. However, the accuracy quickly declines with increasing tilt angle. While a detailed investigation is outside the scope of this study, preliminary tests show that the accuracy of this approach is further limited to thicknesses of at most $100~\mbox{\AA}$. A more fundamental limitation is that the size and shape of any simulation box fulfilling periodic boundary conditions depend on the crystal orientation, specifically the corresponding zone axis $[uvw]$. This complicates the description of a consistent crystal structure with deviations from the ideal periodic structure, generally necessitating a statistical rather than a discrete approach. Localized features such as a discrete twin boundary or a specific crystal morphology can be described more easily in a simulation box with a fixed size. Therefore, the large simulation box in combination with a windowing function was chosen as the most suitable approach for a general simulation pipeline.

\subsection{Simulation parameters and convergence checks} \label{sim_par}

\begin{figure}[!h] %
	\begin{center}
		\includegraphics[width=0.65\textwidth]{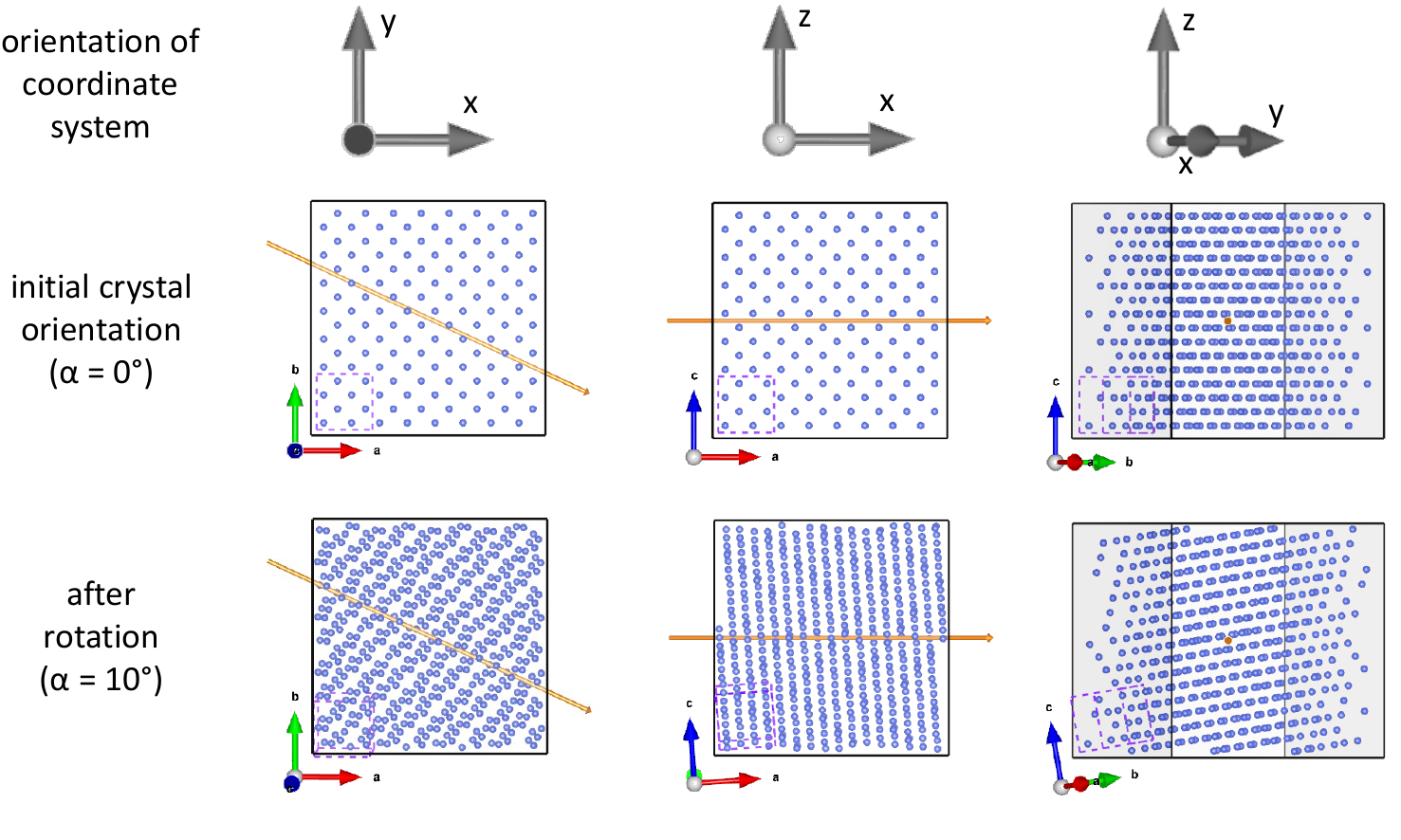}
	\end{center}
	\caption{Orientation of the simulation box with respect to the Cartesian basis ($e_{\mathbf{x}}$, $e_{\mathbf{y}}$, $e_{\mathbf{z}}$). The shown simulation boxes (black edges) have a size of $22 \times 22 \times 22~\mbox{\AA}^3$ and include a model of cubic silicon. Three different viewing directions are depicted for two crystal orientations. The orientation of the crystallographic basis ($\mathbf{a}$, $\mathbf{b}$, $\mathbf{c}$) is shown next to each simulation box. The edges of the unit cell are indicated by dashed, purple lines. The rotation axis is shown as a long orange arrow. The plane wave propagates along the $z$-axis in the positive direction, i.e., towards the viewer.} 
	\label{fig:Orientation}
\end{figure}

Unless stated otherwise, for all simulations in this work, a plane wave with a relativistic electron wavelength of $0.02508~\mbox{\AA}$ was used. This corresponds to a parallel beam of electrons with a kinetic energy of $200~\mbox{keV}$. The plane wave propagates along the Cartesian $z$-axis in the positive direction from the virtual electron source towards the virtual detector. The right-handed crystallographic basis ($\mathbf{a}$, $\mathbf{b}$, $\mathbf{c}$) of all simulated structures was initially oriented such that $a$ points towards the positive direction of $x$. If the unit cell angle $\gamma$ is $90^\circ$, $b$ points exactly in the positive $y$ direction, otherwise the angle between $b$ and positive $y$ is $\gamma - 90^\circ$. The angle between the simulated goniometer rotation axis and the $a$-axis, here called $\omega$, is set to 25$^\circ$ for all cases unless stated otherwise. The geometry and orientation of this simulation setup is shown in Fig.~\ref{fig:Orientation}. 

For initial tests and analysis of the convergence behavior of simulation parameters, coarse-step 3D ED data sets of cubic silicon (diamond structure, $a = 5.43053~ \mbox{\AA}$, space group $Fd\overline{3}m$) were simulated. 16 orientations were defined by rotating the crystal about the $x$-axis from $\alpha = 0^\circ$ to $\alpha = 45^\circ$ in steps of $\Delta \alpha = 3^\circ$. In all cases, the maximum thickness was $1000~\mbox{\AA}$, and diffracted intensities were compared for at least 20 propagation depth values $z$. Only those reflections with $g \leq 2~\mbox{\AA}^{-1}$ were used to assess the convergence. This resolution limit was applied to avoid the contribution of the large number of very weak reflections from the higher-resolution shells and to focus on those reflections that are expected to be measured in a typical 3D ED experiment. For a chosen reference data set, the convergence of each tested parameter \emph{p} was then assessed by calculating the root mean square deviation (RMSD) divided by the root mean square (RMS) (Eq.~\ref{RL2palpha}) of a complete dataset based on all reflection intensity values (without $000$) of all orientations $\alpha$ and thicknesses $z$. In other words, it is the RMSD based on the differences normalised by the RMS of the intensities, which is identical to the Euclidean norm ($L^2$ norm) of the differences divided by the Euclidean norm of the intensities. This is similar to the $wR_{2}$ typically used in crystallographic least-squares refinements if all weights are equal to $1$. Here, we abbreviate this quantity $\textrm{RL2}$. As a criterion to assess if convergence was achieved, we define that $\textrm{RL2}$ must fall below $0.01$. This threshold is marked as a gray line in the subsequent figures. We then define $\textrm{RL2}_{p,\alpha}$ to assess the convergence for one orientation. $\textrm{RL2}_{p}$ is used to assess the convergence based on integrated intensities.

\begin{equation} \label{RL2palpha}
\textrm{RL2}_{p, \alpha} = \sqrt{ \frac{ \frac{1}{N_z N_{\mathbf{h}} }\sum_z \sum_{\mathbf{h}} [I_{\mathbf{h}}(p, z,\alpha) - I_{\mathbf{h}}(p_0, z,\alpha)]^2}{ \frac{1}{N_z N_{\mathbf{h}} }\sum_z \sum_{\mathbf{h}} I_{\mathbf{h}}(p_0,z,\alpha)^2 } }
\end{equation}

\begin{equation} \label{RL2p}
\textrm{RL2}_{p} = \sqrt{ \frac{ \frac{1}{N_z N_{\alpha} N_{\mathbf{h}} }\sum_z \sum_\alpha \sum_{\mathbf{h}} [I_{\mathbf{h}}(p, \alpha, z) - I_{\mathbf{h}}(p_0, \alpha,z)]^2}{ \frac{1}{N_z N_{\alpha} N_{\mathbf{h}} }\sum_z \sum_\alpha \sum_{\mathbf{h}} I_{\mathbf{h}}(p_0,\alpha,z)^2 } }
\end{equation}

\subsubsection{Convergence of Bloch wave parameters}

For convergence tests of the Bloch wave calculations, two parameters controlling the selection of reflections that are used in the structure matrix \textbf{A} were investigated, namely $g_{\mathrm{max}}$ and $S_{g_{\mathrm{max}}}$. To investigate the convergence of $g_{\mathrm{max}}$, an $S_{g_{\mathrm{max}}}$ of $1.0~\mbox{\AA}^{-1}$ was used. To investigate the convergence of $S_{g_{\mathrm{max}}}$, a $g_{\mathrm{max}}$ of $8.0~\mbox{\AA}^{-1}$ was used.

\begin{figure}[!h] % gmax.png
	\begin{center}
		\includegraphics[width=0.75\textwidth]{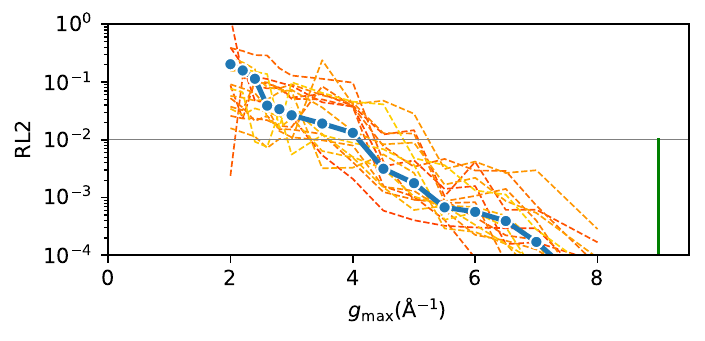} %.eps
	\end{center}
	\caption{$\textrm{RL2}_{p}$ (blue, solid line) and $\textrm{RL2}_{p,\alpha}$ (dashed lines) as a function of Bloch wave parameter $p = g_{\mathrm{max}}$. Darker, more orange colors of $\textrm{RL2}_{p,\alpha}$ indicate larger rotation angles $\alpha$. The calculation with $g_{\mathrm{max}} = 9.0~\mbox{\AA}^{-1}$ (green vertical bar) served as the reference data set.} 
	\label{fig:gmax}
\end{figure}

\begin{figure}[!h] %
	\begin{center}
		\includegraphics[width=0.75\textwidth]{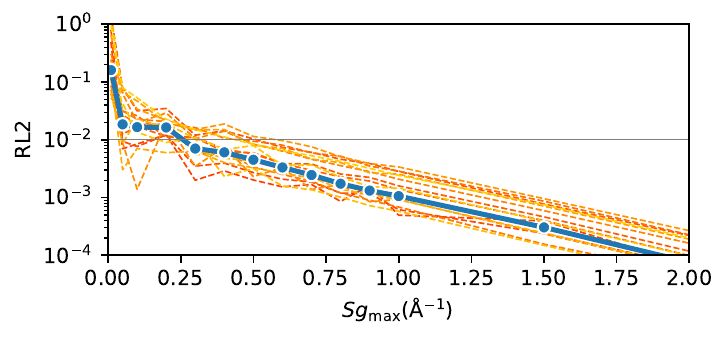} %.eps
	\end{center}
	\caption{$\textrm{RL2}_{p}$ (blue, solid line) and $\textrm{RL2}_{p,\alpha}$ (dashed lines) as a function of Bloch wave parameter $p = S_{g_{\mathrm{max}}}$. Darker, more orange colors of $\textrm{RL2}_{p,\alpha}$ indicate larger rotation angles $\alpha$. The calculation with $S_{g_{\mathrm{max}}} = 4.0~\mbox{\AA}^{-1}$ served as the reference data set.} 
	\label{fig:Sgmax}
\end{figure}

Bloch wave calculations converge rather slowly with $g_{\mathrm{max}}$ (Fig.~\ref{fig:gmax}). Hence, Bloch waves associated with high-resolution shells contribute significantly to the low-resolution shells. To have a consistent set of reflections for comparison, the minimum $g_{\mathrm{max}}$ used was $2.0~\mbox{\AA}^{-1}$. Up to $g_{\mathrm{max}} = 4.0~\mbox{\AA}^{-1}$, the RL2 varies strongly for different orientations. $RL2$ then converges smoothly at $g_{\mathrm{max}} = 6.0~\mbox{\AA}^{-1}$ for all orientations. $S_{g_{\mathrm{max}}}$, in contrast, has in general a weaker effect on calculated intensities and for most orientations representative intensities are obtained with rather small $S_{g_{\mathrm{max}}}$ below $0.2~\mbox{\AA}^{-1}$. For larger values of $S_{g_{\mathrm{max}}}$, the results converge smoothly.

\subsubsection{Convergence of multislice parameters}

\begin{figure}[!h] % Supercell.png
	\begin{center}
		\includegraphics[width=0.75\textwidth]{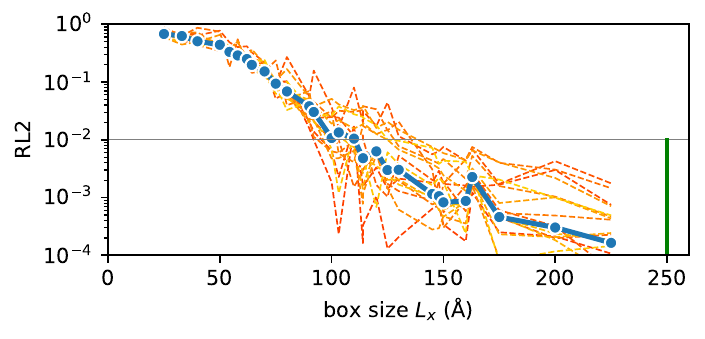}
	\end{center}
	\caption{$\textrm{RL2}_{p}$ (blue, solid line) and $\textrm{RL2}_{p,\alpha}$ (dashed lines) as a function of the box size parameter $L_{x} = L_{y}$. Darker, more orange colors of $\textrm{RL2}_{p,\alpha}$ indicate larger rotation angles $\alpha$. The reference data set is based on $L_{x} = 250~\mbox{\AA}$ (green vertical bar).} 
	\label{fig:Supercell}
\end{figure}

Besides standard sampling and slicing parameters, the convergence of the simulation box size needed to be checked. For this purpose, the size $L_{x} = L_{y}$ was varied in the range between $25~\mbox{\AA}$ and $250~\mbox{\AA}$ (Fig.~\ref{fig:Supercell}). A sampling $s_{x} = s_{y} = 0.04~\mbox{\AA}$ and slice thickness $s_{z} = 0.8~\mbox{\AA}$ were used. For $L_{x} < 100~\mbox{\AA}$, the $\textrm{RL2}$ still shows a strong dependence on $L_{x}$. Above $100~\mbox{\AA}$, calculations approach convergence, though there are cases where a slight increase of the box size leads to a jump of $\textrm{RL2}$. In combination with the dependence on the orientation angle, we consider a supercell size of $160~\mbox{\AA}$ to be sufficient. For the silicon structure, this corresponds to about 30 times the unit cell parameter $a$. The choice of $160~\mbox{\AA}$ was also checked for compounds with larger unit cell volumes and considered a suitable value. Later it will be shown, that a larger supercell size may be required for larger thicknesss for cases with increased interaction strength, i.e., for high-Z compounds or smaller electron kinetic energies (SI Section~\ref{SI_interaction_strength}).

\begin{figure}[!h] % Slice\_thickness.png
	\begin{center}
		\includegraphics[width=0.75\textwidth]{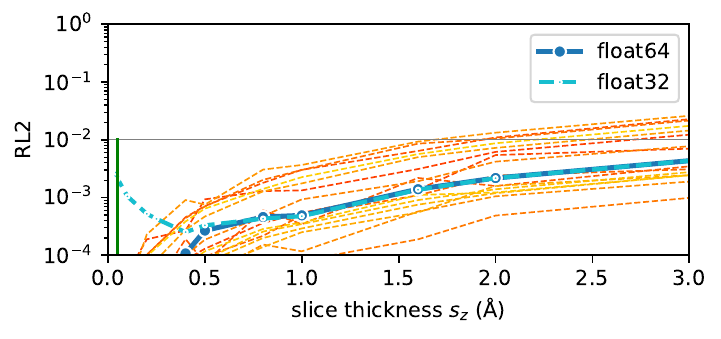}
	\end{center}
	\caption{$\textrm{RL2}_{p}$ as a function of the slice thickness $s_{z}$ and numerical precision used. For double-precision ("float64", solid line), the $\textrm{RL2}_{p,\alpha}$ (dashed lines) for individual orientations are plotted where darker, more orange colors indicate larger rotation angles $\alpha$. The double-precision calculations with $s_{z} = 0.05~\mbox{\AA}$ (green vertical bar) served as the reference data set.} 
	\label{fig:Slice_thickness}
\end{figure}

The electrostatic potential of the atoms in the box is sliced parallel to the beam propagation direction in steps of $s_{z}$, representing the slice thickness. Each slice is projected onto the $xy$ plane so that slices become a 2D potential which is sampled in steps of $s_{x} = s_{y}$. The slice thickness was tested in the range from $0.1~\mbox{\AA}$ to $4~\mbox{\AA}$. To achieve a consistent set of simulated crystal thicknesses from calculations with different slice thicknesses, comparisons only used a subset of the data where the thickness is an integer multiple of $16~\mbox{\AA}$ (Fig.~\ref{fig:Slice_thickness}). Another parameter related to the slicing is the projection method. It is computationally faster to assign the entire electrostatic potential of an independent atom to a single slice. This is called the infinite projection method. The more precise approach is to slice an atom's electrostatic potential according to the actual slice boundaries, which is called the finite projection method. Preliminary tests indicate that this leads to a negligible improvement regarding the agreement with Bloch wave calculations (SI Section~\ref{Projection_method}). Furthermore, calculations with the finite projection approach take about 15 times longer than those with the simpler infinite projection. Independent of the projection approach, the numerical precision plays a crucial role. For slice thickness values below $0.4~\mbox{\AA}$, the calculations begin to diverge if they are based on single-precision floating-point numbers ("float32"). Calculations with double-precision floating-point numbers ("float64") converge smoothly and do not diverge. As the latter extends the calculation times significantly by a factor of about 5 (SI Fig.~\ref{fig:Walltime} ), single-precision calculations with $s_{z}$ between $0.4$ and $0.5~\mbox{\AA}$ and infinite projection are used for all subsequent calculations.

\begin{figure}[!h] % Sampling.png
	\begin{center}
		\includegraphics[width=0.75\textwidth]{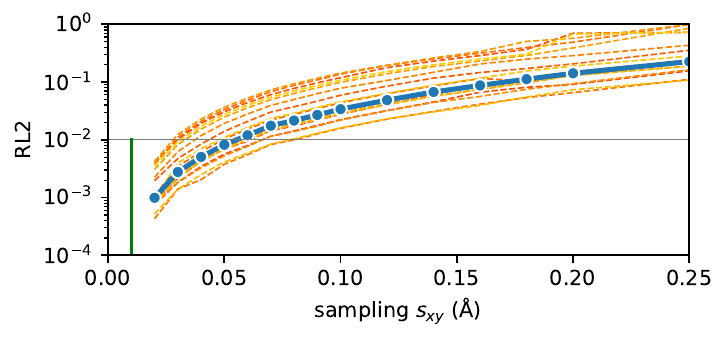}
	\end{center}
	\caption{$\textrm{RL2}_{p}$ (solid line) as a function of the sampling $s_{x} = s_{y}$. $\textrm{RL2}_{p,\alpha}$ (dashed lines) for individual orientations are plotted where darker, more orange colors indicate larger rotation angles $\alpha$. The double-precision calculations with $s_{x} = 0.01~\mbox{\AA}$ (green vertical bar) served as the reference data set.} 
	\label{fig:Sampling}
\end{figure}

The real-space sampling of the slices defines the number of gridpoints $N = (L_{x}/s_{x})(L_{y}/s_{y})$ of the 2D slices, i.e. of the electrostatic potential and of the propagating wavefunction. Due to the time complexity of $O(N \log N)$ of the fast Fourier transform algorithm, the sampling is the parameter with the strongest influence on the simulation runtime and it is thus the bottleneck of the simulation of 3D ED datasets with multislice. The convergence of the sampling was investigated in a series of simulations with $s_{x} = s_{y}$ ranging from $0.01~\mbox{\AA}$ to $0.64~\mbox{\AA}$ (Fig.~\ref{fig:Sampling}). The $\textrm{RL2}$ shows a smooth dependence on $s_{x}$, though full convergence possibly requires an even finer sampling with $s_{x} < 0.01~\mbox{\AA}$. However, the associated computational cost and memory requirements increase significantly with finer sampling. For this study, we choose a sampling of $0.04~\mbox{\AA}$ as a balance between accuracy and computational cost.

\subsection{Comparisons of results from Bloch wave and multislice calculations} \label{sec:Comparisons}

Diffracted intensities from Bloch wave and multislice calculations are compared using reflections with $g_{\mathbf{h}} \leq 2.0~\mbox{\AA}^{-1}$. The application of the Hann window $w(x,y)$ in the multislice approach results in a nominal scaling of amplitudes in Fourier space by a factor $f_{s} = \left< w(x,y) \right> = \frac{3}{8}$, where $\left< \right>$ denotes the spatial mean of the window function. Consequently, diffracted intensities are scaled by $f_{s}^2 = \left( \frac{3}{8} \right)^2$. A comparison of the two approaches thus requires the multiplication of diffracted intensities determined with multislice by $1/f_{s}^{2}$. The agreement between intensities determined with the two approaches is then assessed via residual factors \emph{R}. Two different \emph{R}-factors are used in this work. First, $R_{z,\alpha}$ is related to intensities from one diffraction pattern corresponding to a certain crystal thickness and orientation. For each data set, there are $N_{z} \times N_{\alpha}$ of these single-frame $R_{z,\alpha}$-factors (Equ.~\ref{R_zalpha}). By summing up the intensities of one reflection across different crystal orientations, integrated intensities $I_{\mathbf{h}}(z)$ are determined, here neglecting any Lorentz or other geometry-related corrections. \emph{R}-factors based on integrated intensities are labelled $R_{z}$ and correspond to the Bragg \emph{R}-factor typically used in crystal structure refinement. For each data set, there are $N_{z}$ of these $R_{z}$-factors (Equ.~\ref{R_z}).

\begin{equation} \label{R_zalpha}
R_{z, \alpha} = \frac{\sum_\mathbf{h}|{I}_\mathbf{h}^{\mathrm{BW}}(z, \alpha) - {I}_\mathbf{h}^{\mathrm{MS}}(z, \alpha)|}{\sum_\mathbf{h}{I}_\mathbf{h}^{\mathrm{BW}}(z, \alpha)} 
\end{equation}

\begin{equation} \label{R_z}
R_{z} = \frac{\sum_\mathbf{h}\sum_{\alpha}|{I}_\mathbf{h}^{\mathrm{BW}}(z,\alpha) - {I}_\mathbf{h}^{\mathrm{MS}}(z,\alpha)|}{\sum_\mathbf{h}\sum_{\alpha}{I}_\mathbf{h}^{\mathrm{BW}}(z,\alpha)}
\end{equation}

\section{Results}

\begin{figure}[!h] % Rocking\_curves.png
	\begin{center}
		\includegraphics[width=0.75\textwidth]{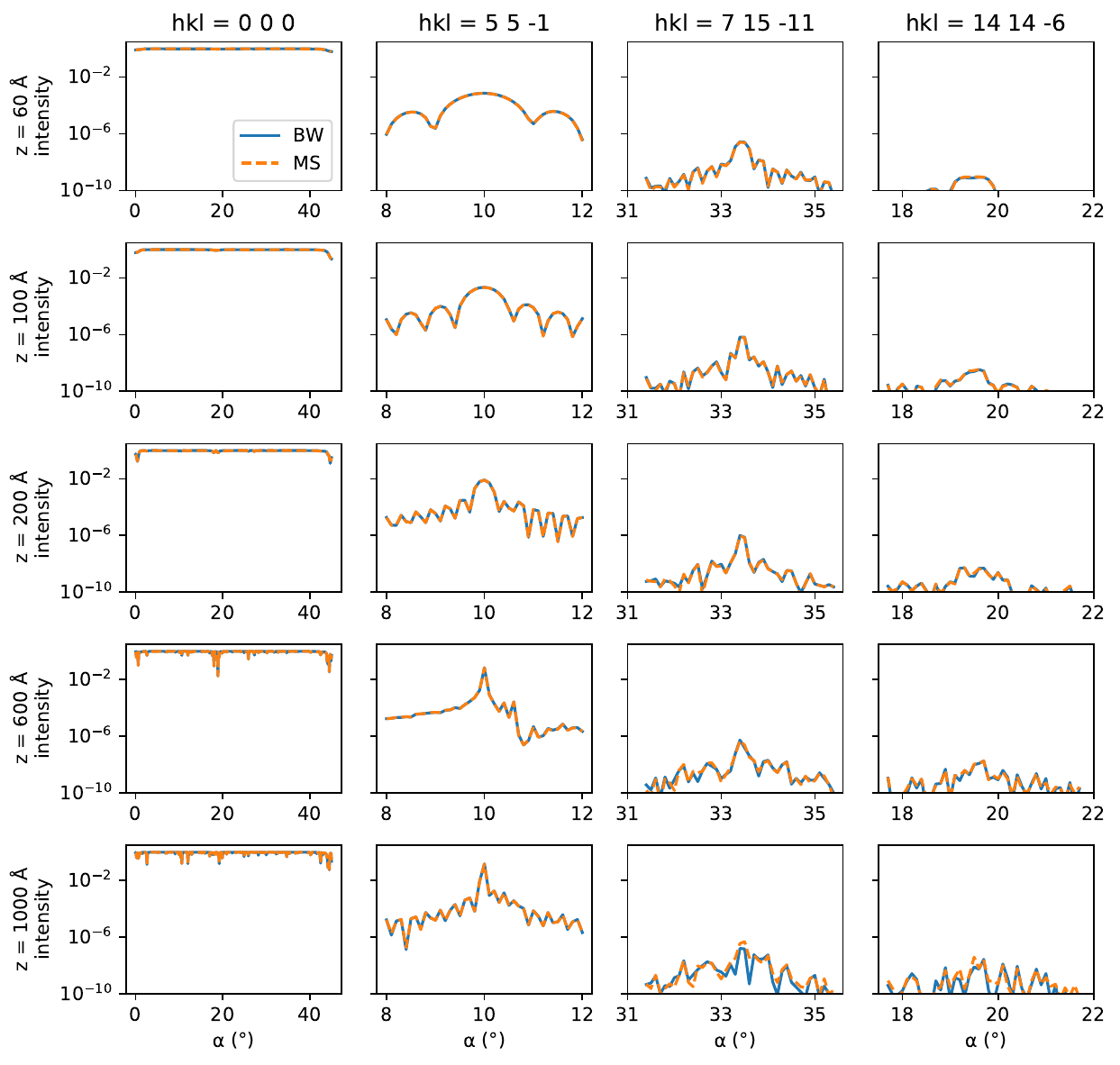}
	\end{center}
	\caption{Selected rocking curves (reflection profiles) of Si at five different crystal thicknesses. The intensities obtained with the Bloch wave (BW) approach are shown as solid lines, while those obtained with the multislice (MS) approach are shown as dashed lines. The first column shows the intensity of the $000$ reflection for the entire angular range for $\alpha$ from $0^\circ$ to $+45^\circ$. In the other columns, the angular range only covers the region close to the maximum intensity of the respective reflection. In this particular case, $\omega$ was set to $0^\circ$.}
	\label{fig:Rocking_curves}
\end{figure}

\subsection{Agreement between simulated 3D ED data sets based on Bloch wave and multislice approaches} \label{sim_Si}

Various 3D ED data sets of silicon were simulated as a series of 451 orientations by rotating Si-based crystal structures about the $x$-axis, as defined in Fig.~\ref{fig:Orientation} by an angle $\alpha$ from $0^\circ$ to $+45^\circ$ in steps of $\Delta \alpha = 0.1^\circ$. Various orientations of the rotation axis were used given as the angle $\omega$. Several parameters, including unit-cell parameters, atomic number of atoms in the unit cell, and the kinetic energy of electrons were systematically varied and their effect on diffracted intensities investigated. Data sets were generated using both the Bloch wave approach as well as the multislice method. Figure~\ref{fig:Rocking_curves} shows selected rocking curves of cubic Si at different crystal thicknesses. The results demonstrate a very good agreement between the two approaches, with only minor deviations observed for weaker reflections. Deviations are more pronounced for larger thicknesses, probably related to the accumulation of numerical noise. The overall excellent agreement between the two approaches is considered a further assessment of the convergence of the chosen simulation parameters.

\subsubsection{Dependence on thickness}

\begin{figure}[!h] %
	\begin{center}
		\includegraphics[width=0.75\textwidth]{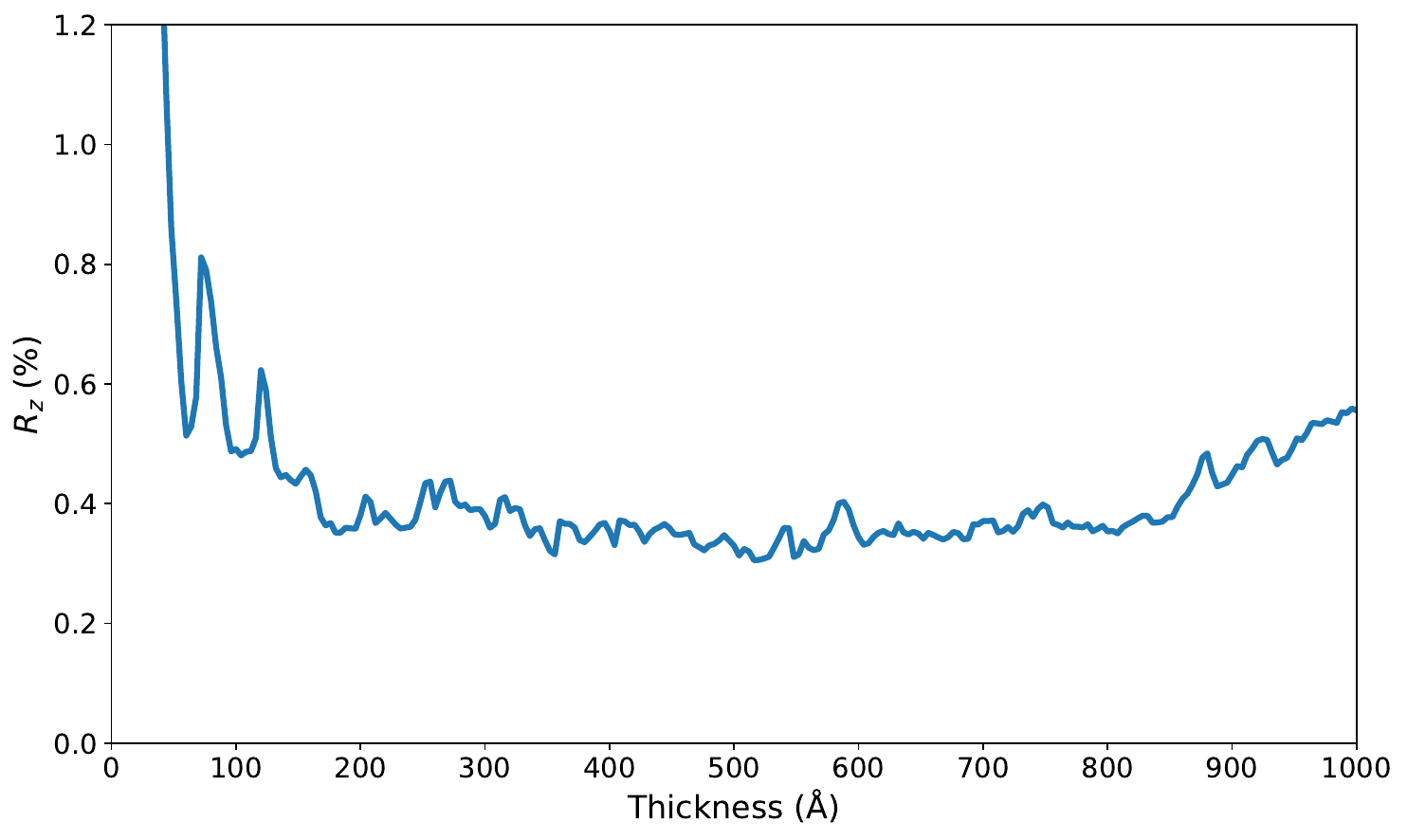}
	\end{center}
	\caption{$R_{z}$ based on intensities obtained with Bloch wave and multislice approaches. In this particular case, $\omega$ was set to $0^\circ$.}
	\label{fig:R_vs_z}
\end{figure}

The $R_{z}$ as a function of crystal thickness was used to quantify the agreement between the diffraction intensities obtained with the Bloch wave and multislice approaches. The results presented in Fig.~\ref{fig:R_vs_z} reveal a distinct trend: at very small thicknesses, $R_{z}$ starts at relatively high values. While the Bloch wave approach is based on the scattering power of complete unit cells, multislice is affected by the upper boundary of the simulation box representing the crystal surface. For very thin crystals, surface effects have a stronger influence on the diffracted intensities determined with multislice and thus on the agreement with Bloch wave calculations. As the crystal thickness increases, the bulk volume, which is treated consistently by both the Bloch wave and multislice methods, begins to dominate, leading to a very good agreement. With further increasing thickness, $R_{z}$ slightly increases which is attributed to parameter choices, especially sampling and simulation box size. Subtle methodological discrepancies and numerical noise inevitably accumulate over extended propagation distances, causing $R_{z}$ to gradually increase for thicker crystals. This is further investigated in SI Section~\ref{SI_interaction_strength}.

\subsubsection{Dependence on kinetic energy}

\begin{figure}[!h] %
	\begin{center}
		\includegraphics[width=0.75\textwidth]{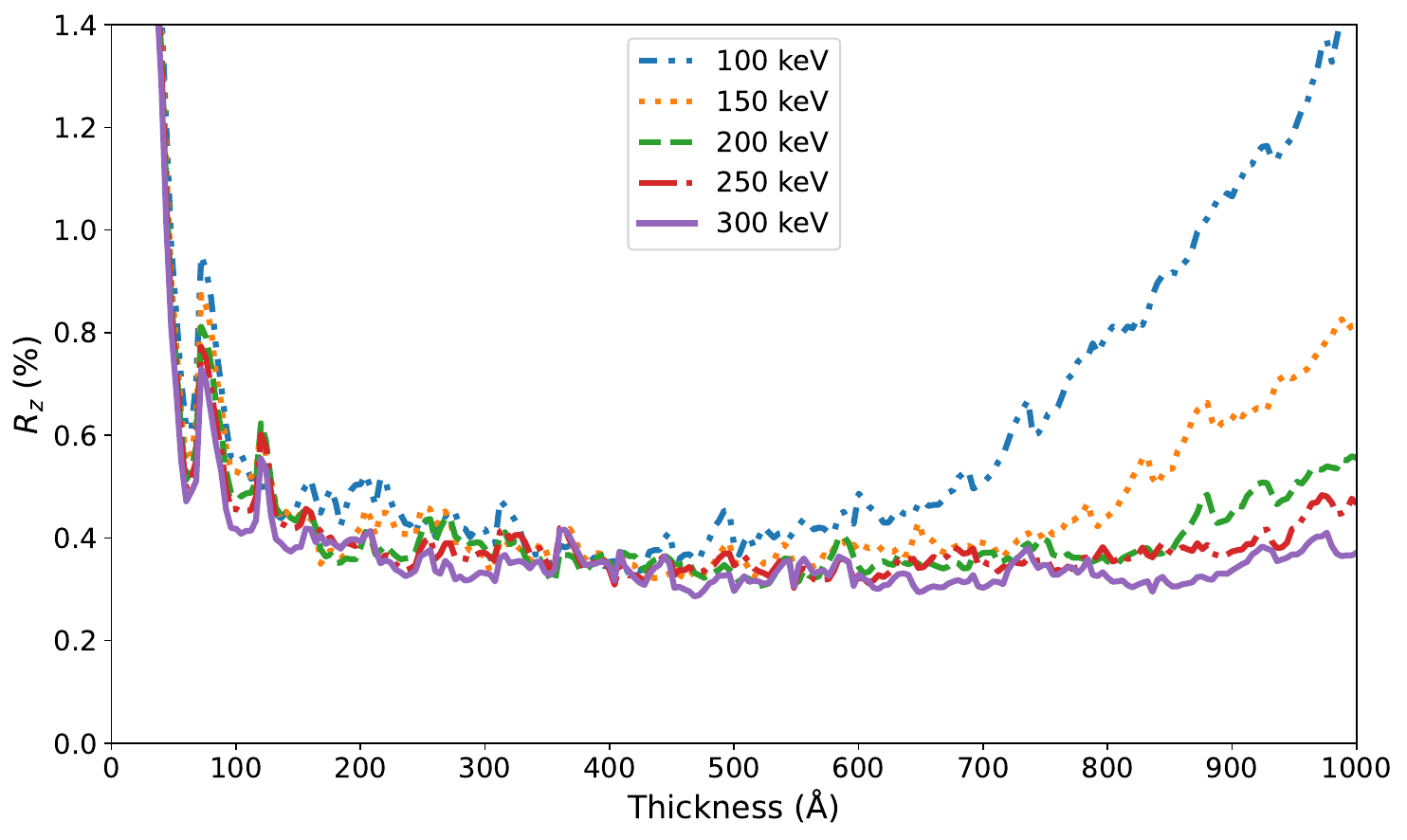}
	\end{center}
	\caption{${R_z}$ as a function of kinetic energy and sample thickness.} 
	\label{fig:Energy}
\end{figure}

The kinetic energy directly influences the electron wavelength and the interaction strength. A series of simulations was performed for kinetic energies in the range from $100$ to $300~\mbox{keV}$ (Fig.~\ref{fig:Energy}). As energy increases and thus the interaction strength decreases, $R_{z}$ tends to decrease. A further analysis revealed that a stronger interaction requires a larger simulation box (SI, Fig.~\ref{fig:R_vs_keV}) as deviations arising from not fully converged parameters and numerical noise are amplified.

\subsubsection{Dependence on orientation of rotation axis}

To further assess the agreement between the Bloch wave and multislice methods, we tested different crystal orientations relative to the rotation axis. Four initial crystal orientations were tested, each with a defined angle $\omega$ between the rotation axis and the $[100]$ direction Fig.~\ref{fig:Orientation}. For $\omega = 0^\circ$, the simulation box fulfills periodic boundary conditions along $y$ for all rotation angles $\alpha$. The $R_{z,\alpha}$ map (Fig~\ref{fig:Orientation}) shows a periodic pattern of thicknesses defining stripes of high-\emph{R}. This pattern arises from because for all orientations the $400$ and $\bar{4}00$ reflection are the strongest reflections and their interplay with the 000 dominate the intensity distribution. The periodicity of the pattern corresponds to the extinction distance, which is identical for the two reflections. Thickness ranges with high \emph{R} correspond to thicknesses where the $400$ and $\bar{4}00$ reflections are weak and the 000 reflection is strong. As this special conditions is fulfilled for all $\alpha$ for this special orientation of the rotation axis, the agreement is dominated by a small number of reflections, amplifying numerical differences between the Bloch wave and multislice methods. This is not representative for the typical 3D ED experiment. The other simulations with $\omega$ set to $5^\circ$, $13^\circ$ or $25^\circ$ avoid low-index zone axes for all simulation parameters and compounds in this study except for the $[001]$ zone axis if $\alpha = 0^\circ$. Without a clear preference, an orientation of $\omega = 25^\circ$ was selected as the starting point for all subsequent simulations.

\begin{figure}[!h] %
	\begin{center}
		\includegraphics[width=0.95\textwidth]{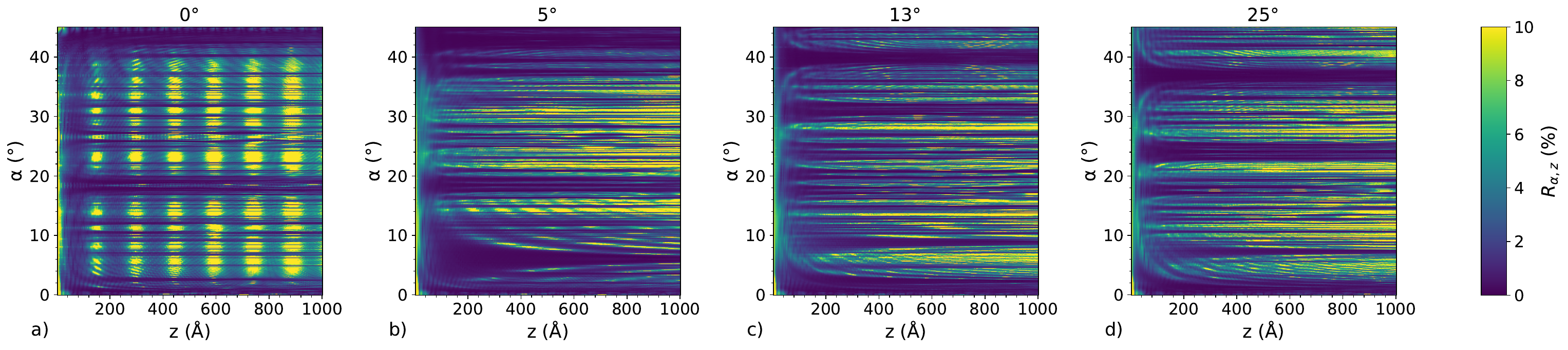}
	\end{center}
	\caption{$R_{z,\alpha}$ as a function of starting orientation of the crystal relative to the rotation axis. a) $0^\circ$, b) $5^\circ$, c) $13^\circ$ and d) $25^\circ$.} 
	\label{fig:Orientation2}
\end{figure}

\subsubsection{Dependence on symmetry}

\begin{figure}[ht] %
	\begin{center}
		\includegraphics[width=0.75\textwidth]{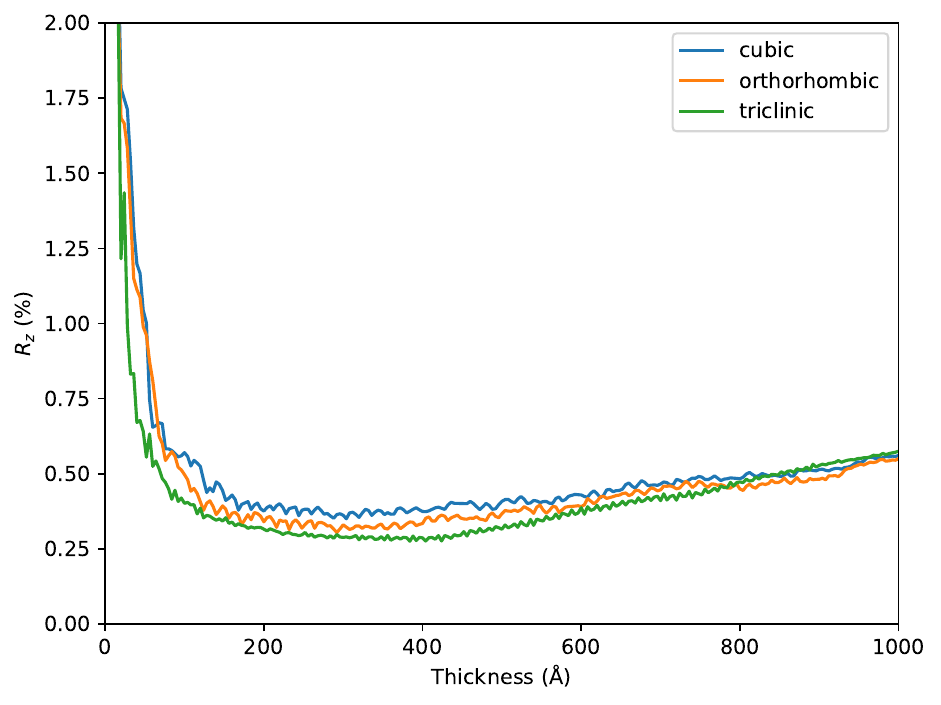}
	\end{center}
	\caption{$R_{z}$ of cubic silicon (blue), distorted silicon with orthorhombic unit cell metric (orange), and distorted silicon with triclinic unit cell metric (green). For all three cases, the unit cell volume and the density is identical. Here, $\omega = 25^\circ$.}
	\label{fig:Symmetry} 
\end{figure}

To evaluate the suitability of our simulation pipeline across different crystal symmetries, distorted variants of the Si structure with identical unit cell volumes were generated. First, an orthorhombic variant was created by setting the lattice parameters to $5.20305$, $5.7$, and $5.4~\mbox{\AA}$. Second, a triclinic variant was generated with unit cell parameters of $5.3$, $5.8$, and $5.4~\mbox{\AA}$, and unit cell angles of $80^\circ$, $95.4^\circ$, and $101^\circ$. Figure~\ref{fig:Symmetry} shows clearly that the differences in $R_{z}$ are negligible and there is no distinct pattern to be observed to suggest that there is an influence of particular symmetry on the convergence of both methods.

\subsubsection{Dependence on atomic number \emph{Z}}

\begin{figure}[!h] %
	\begin{center}
		\includegraphics[width=0.75\textwidth]{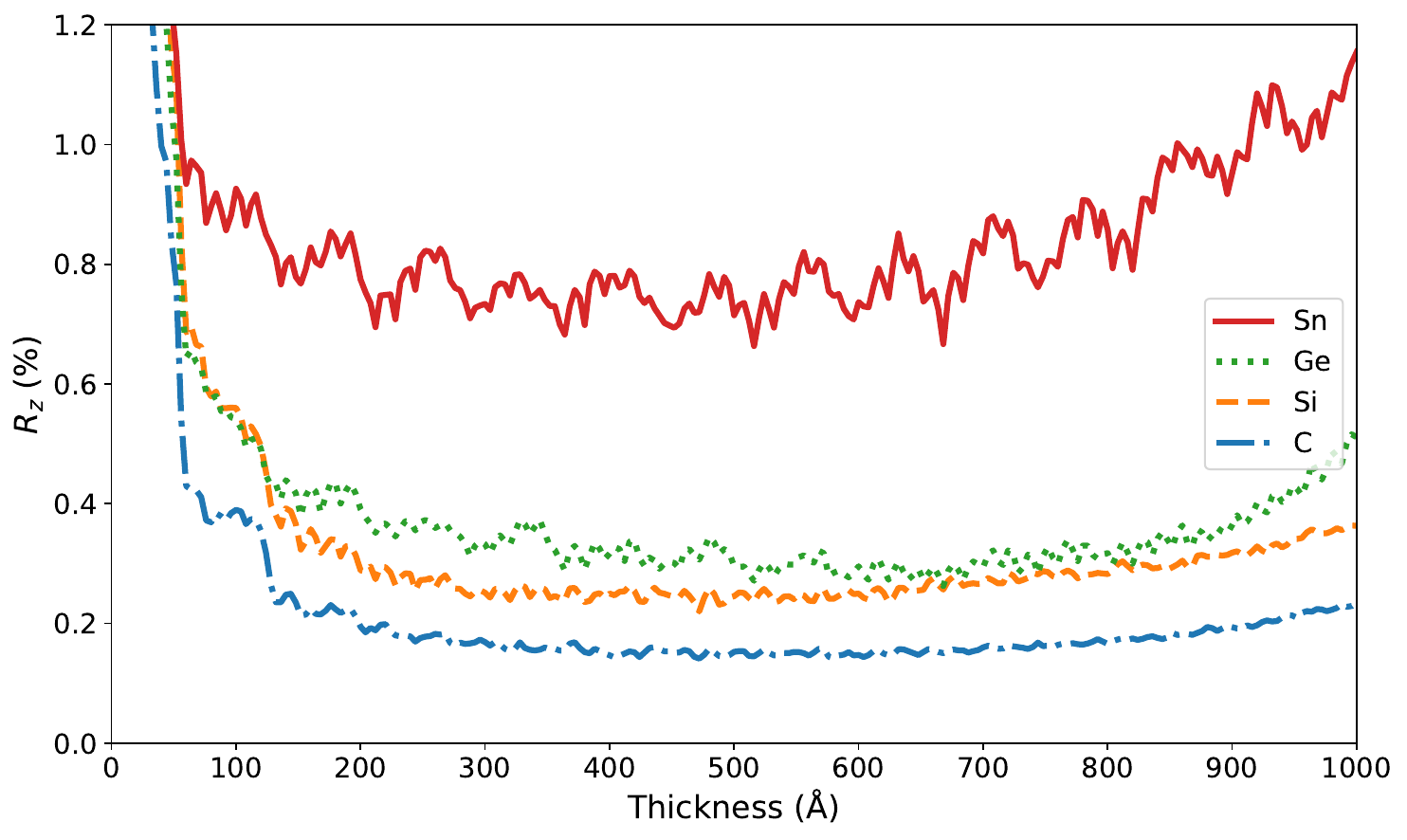}
	\end{center}
	\caption{Changes in \emph{R}-factor according to different atoms substituting Si in the unit cell.} 
	\label{fig:R_vs_atoms}
\end{figure}

To isolate the effect of atomic number ($Z$) on simulations, the silicon site in the was systematically substituted with other group IV elements (C, Ge, and Sn). Maintaining fixed unit cell geometry and symmetry across all models ensured that observed differences stemmed directly from variations in atomic scattering power rather than other structural factors. Higher $R_{z}$ are observed for the heavier elements with higher electrostatic potential and associated stronger electron scattering power (Fig.~\ref{fig:R_vs_atoms}). In all cases, a slight increase in $R_{z}$ is observed for larger thicknesses. This increase is more pronounced for heavier atoms, which alignes with the trends already observed before: conditions enhancing the interaction strength require computationally more expensive simulation parameters (SI Section \ref{SI_interaction_strength}). 

\subsection{Realistic example cases}

\begin{figure}[!h]
	\begin{center}
		\includegraphics[width=0.95\textwidth]{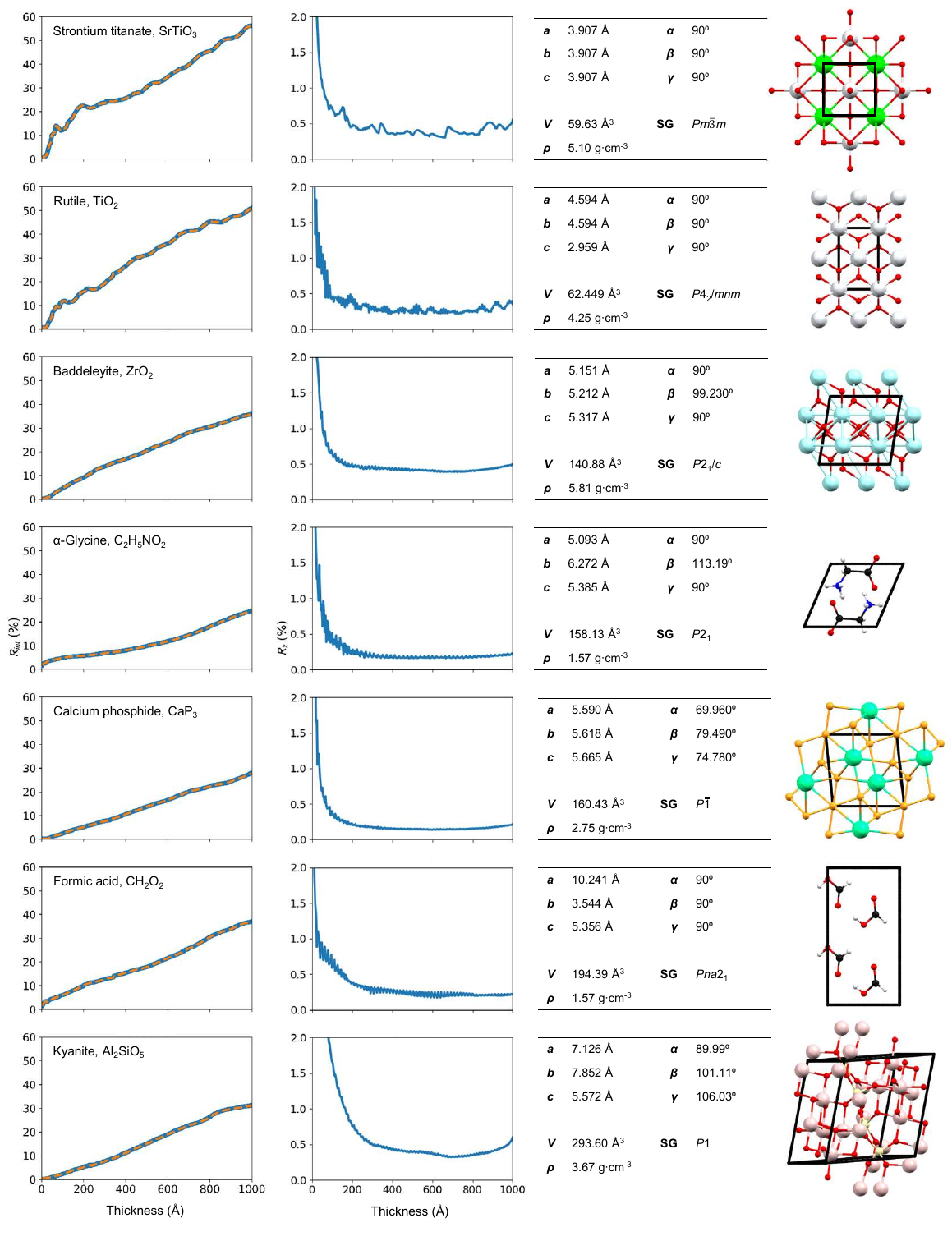} 
	\end{center}
	\caption{Simulation results for the seven selected crystal structures, encompassing variety in elemental composition, unit cell sizes and symmetry of the structures.} 
	\label{fig:Realistic_cases}
\end{figure}

To demonstrate the generality and robustness of our simulation approach, we extended our analysis from benchmark silicon-based structures to a set of simple crystalline compounds covering a range of symmetries and chemical compositions. This section presents the simulation results for $\mathrm{CaP_{3}}$, $\mathrm{Al_{2}SiO_{5}}$, $\mathrm{SrTiO_{3}}$, $\mathrm{TiO_{2}}$, $\mathrm{ZrO_{2}}$, and two organic compounds, glycine ($\mathrm{C_{2}H_{5}NO_{2}}$) and formic acid (HCOOH). We chose structures with various symmetries: triclinic $\mathrm{CaP_{3}}$ and $\mathrm{Al_{2}SiO_{5}}$, monoclinic glycine and $\mathrm{ZrO_{2}}$, tetragonal $\mathrm{TiO_{2}}$, orthorhombic formic acid, and cubic $\mathrm{SrTiO_{3}}$. We also chose centrosymmetric and non-centrosymmetric crystals, and picked structures with more than one type of atom in the unit cell. In all cases, the same crystal model orientations are used as defined before, i.e., the angle $\omega$ between $\mathbf{a}$ and $\mathbf{x}$ is $25^\circ$, and a sample holder rotation about $\mathbf{x}$ is simulated for the $\alpha$ range between $0^\circ$ and $45^\circ$ in steps of $0.1^\circ$. The Bloch wave approach and the multislice method are used and compared. For both approaches, the internal \emph{R}-factor $R_{\mathrm{int}}$ is calculated as a function of thickness. The calculation of $R_{\mathrm{int}}$ was based on reflections, for which a complete rocking curve was determined and for which at least one other symmetrically equivalent reflection was in the data set. Details are given in the Supplementary Information. This quantity expresses how similar intensities of symmetrically equivalent reflections are. In all cases, it is close to 0\% for small thicknesses and then increases steadily with thickness (Fig.~\ref{fig:Realistic_cases}). As the number of reflections with complete rocking curves changes as a function of thickness, the curves are not monotonic. With this quantity, the simulations assess the suitability of the kinematic approximation, which is better fulfilled for very low thicknesses. The $R_{\mathrm{int}}$ based on multislice and on Bloch wave calculations are essentially identical, demonstrating that both approaches are suitable for the simulation of 3D ED data sets. Independent of this, the agreement between the two simulation formalisms was further quantified by $R_{z}$ (Fig.~\ref{fig:Realistic_cases}, SI Section \ref{Realistic_cases}).

The determined $R_{\mathrm{int}}$ values across this diverse set of compounds reveal trends consistent with expectations and test calculations. First, an increase in crystal thickness steadily leads to a higher $R_{\mathrm{int}}$. This is in line with the expectation that thicker crystals exhibit more pronounced dynamical effects, which translates to greater deviations from kinematic intensities. Second, higher material density consistently leads to a higher $R_{\mathrm{int}}$ as a stronger interaction between the electron beam and the electrostatic potential results in more pronounced dynamical effects.

%{Figures/Realistic\_cases\_full.png}

\section{Summary and Outlook}

This work demonstrates the simulation of continuous-rotation 3D ED data. For this purpose, a versatile pipeline was implemented based on the \emph{ab}TEM package. Two independent simulation branches based on the Bloch wave formalism and the multislice method were pursued. By rigorously cross-checking these two approaches, we demonstrated an unprecedented agreement between them for high- and low-symmetry compounds with up to four different elements in the unit cell. Establishing this consistency required a thorough convergence check of key simulation parameters for Bloch wave calculations (resolution limit, excitation error) and for the multislice approach (box size, slice thickness, sampling). Due to the violation of periodic boundary conditions for tilted specimen in the multislice approach, the application of a windowing function for the exit wave was essential. We also observed a direct correlation between the strength of the electron-matter interaction and the requirement for larger simulation boxes.

This work establishes guidelines for obtaining physically meaningful and computationally efficient simulation of 3D ED datasets. With an inexpensive desktop computer, the simulation of data sets with the Bloch wave approach typically takes several minutes for the smallest unit cells, and around one hour for larger unit cells presented in this article. In contrast, the multislice approach is computationally more expensive, typically requiring approximately one day of runtime. If the demonstrated level of accuracy is not needed, runtimes can be reduced substantially.

The present model is based on static crystal structures and purely accounts for elastic scattering. Due to the difference in computational costs, the Bloch wave branch is especially suitable for high throughput simulation of training data for, e.g., machine-learning applications \cite{PhAI2024, Malik2026} or statistical analysis of dynamical effects on quantities typically used in crystallographic structure analysis ($R_{\mathrm{int}}$).

There are numerous examples that demonstrate extremely good agreement between measured and calculated data both on the Bloch wave \cite{Zuo1992, Palatinus2015a, Palatinus2015b} and multislice approach \cite{Jansen1998}. However, to the best of our knowledge, multislice has not been used for structure model refinements against experimental 3D ED data. Based on the successful cross-validation of the Bloch wave and multislice branches, our results confirm the general suitability of multislice calculations to contribute to 3D ED.

We see a promising foundation for future extensions and methodological developments to achieve more realistic, overarching simulations. For example, inelastic scattering phenomena \cite{Mendis2025} may be implemented. Bloch wave-based simulations are computationally attractive only for structural features that can be represented in the time- and space-averaged unit cell. For deviations of the periodic structure (e.g., grain boundaries, twin boundaries, mosaicity, thermal displacements), the multislice approach based on a large simulation box allows for the modelling of such structural features. 

Building upon our current work, we will focus on extending the pipeline to address several currently omitted aspects that are crucial for a comprehensive understanding of electron diffraction beyond elastic scattering of idealised defect-free crystals. One strength of the envisioned extended simulation pipeline will be that each of the mentioned structural features can be switched on and off so that their contribution to diffracted intensities can be investigated and assessed. 

Beyond standard 3D ED simulations, the versatility of this framework naturally extends to other advanced electron microscopy techniques. The pipeline can be adapted for the simulation of four-dimensional scanning transmission electron microscopy (4D-STEM) and 4D-STEM tomography.

\clearpage

%%%%%%%%%%%%%%%%%%%%%%%%%%%%%%%%%%%%%%%%%%%%%%%%%%
%%%%%%%%%%%%%%%%%%%%%%%%%%%%%%%%%%%%%%%%%%%%%%%%%%
 
\appendix % if required
\setcounter{figure}{0}
\renewcommand{\thefigure}{A.\arabic{figure}}
\section{Supplementary Information}

\subsection{Computational pipeline} \label{Computational_pipeline}

\subsubsection{Installation}
A Python package called py3DED is available on github (https://github.com/3DED/py3DED), where detailed installation instructions are provided. The version used for this study is from July 2026 and has the commit hash 15cb5f8. It is recommended to install the package in a new, isolated Python environment to prevent potential dependency conflicts. In the repository, Python scripts for running the simulation and analysing output files are provided alongside suitable input files.

\subsubsection{Input files}
The program requires two input files: a JSON configuration file and a CIF file. The JSON file specifies the simulation parameters and includes the path to a CIF file. The CIF file contains the crystal structure information. Note that displacement parameters and occupancies defined in the CIF file are ignored. One displacement parameter per element can be defined in the JSON file, which is then applied to all atoms of that element. The CIF file is read using the \emph{Atomic Simulation Environment} package (ASE) \cite{Larsen2017}, which in some cases may not be able to read certain CIF files. In such cases, the user is advised to convert the CIF file to a format that ASE can read (e.g., VESTA).
	
Here is an exmple input file with parameters that will run Bloch wave and multislice calculations. The output file will have identically the same structure, which facilitates the comparison of the two approaches. 

\begin{verbatim}
	
	{
		"cif_file": "/path/to/crystal_structure.cif",
		"output_folder": "/path/to/output/directory/",
		"device": "gpu",
		"precision": "float32",

		"mode": "bw+ms",
		"energy": 200000,

		"thermal_sigmas": 0.1,
		"centering": "P",

		"rotation_axis_orientation": 25.0,
		"rotation_min": 0.0,
		"rotation_max": 45,
		"rotation_steps": 451,
		
		"thickness": 1000,
		"slice_thickness": 0.5,
		"exit_planes": 10,

		"box_size_x": 160,
		"box_size_y": 160,
		"sampling": 0.04,

		"projection": "infinite",
		"window_func": "hann",
		"integration_radius": 0.02,

		"g_max": 8.0,
		"sg_max": 0.5,
		
		"g_max_store": 2.0,
	}
\end{verbatim}

Here is a description of selected parameters.
		\begin{enumerate}
			\item {\fontfamily{pcr}\selectfont cif\_file}: specifies the path to the Crystallographic Information File (CIF) used as input for the simulation.
			
			\item {\fontfamily{pcr}\selectfont output\_folder}: defines the absolute path to the directory where all generated simulation data and results will be stored.
			
			\item {\fontfamily{pcr}\selectfont device}: specifies the computational device to be used for running the simulations. “gpu” is for Graphics Processing Unit (GPU), “cpu” is for using a normal processor.
			
			\item {\fontfamily{pcr}\selectfont precision}: determines floating-point precision of multislice calculations and of intensity values in output files (for both multislice and Bloch wave).
			
			\item {\fontfamily{pcr}\selectfont mode}: either "bw" (Bloch wave approach), "ms" (multislice approach) or "bw+ms" (both approaches).
			
			\item {\fontfamily{pcr}\selectfont energy}: defines the kinetic energy of the electrons in electron volts (eV).
			
			\item {\fontfamily{pcr}\selectfont thermal\_sigmas}: defines the thermal displacement parameter $\sigma=\sqrt{U_{iso}}$ related to the Debye-Waller factor.
			
			\item {\fontfamily{pcr}\selectfont centering}: Centering of the unit cell, 'P' for primitive, 'I' for body-centered, 'F' for face-centered, or 'C' for base-centered. This parameter ist not automatically determined. It is used for Bloch wave calculations to setup a structure matrix and for the indexing step in multislice.
			
			\item {\fontfamily{pcr}\selectfont rotation\_axis\_orientation}: specifies the angle between the rotation axis and the Cartesian $x$-axis in the $xy$-plane in degrees.
			
			\item {\fontfamily{pcr}\selectfont rotation\_min, rotation\_max}: define the range of orientations to simulate.
			
			\item {\fontfamily{pcr}\selectfont rotation\_steps}: specifies the number of discrete orientations to simulate within the defined range.

			\item {\fontfamily{pcr}\selectfont thickness}: specifies the maximum thickness of the crystal (in \AA) being simulated.
			
			\item {\fontfamily{pcr}\selectfont slice\_thickness}: defines the thickness (in \AA) of each discrete slice into which the crystal is divided along the beam direction for multislice calculations.
			
			\item {\fontfamily{pcr}\selectfont exit\_planes}: number of exit planes or slices for which the diffraction pattern is calculated and used to extract diffracted intensities in multislice simulations.

			\item {\fontfamily{pcr}\selectfont box\_size\_x} and {\fontfamily{pcr}\selectfont box\_size\_y}: define the lateral dimensions (in \AA) of the simulation box used in multislice calculations.
			
			\item {\fontfamily{pcr}\selectfont sampling}: real-space sampling interval (in \AA) used in multislice simulations, which is the discretization of the projected potential and of the wavefunction.

			\item {\fontfamily{pcr}\selectfont projection}: specifies how the electrostatic potential is projected along the beam direction. 

			\item {\fontfamily{pcr}\selectfont window\_func}: specifies the window function to be applied to the wave function at the exit plane in multislice calculations.

			\item {\fontfamily{pcr}\selectfont integration\_radius}: defines the radius (in $\mbox{\AA}^{-1}$) within which the intensity of Bragg reflections is integrated from calculated diffraction patters (only used in multislice calculations).
			
			\item {\fontfamily{pcr}\selectfont g\_max}: specifies the the resolution limit of Bloch wave calculations as the maximum length of reciprocal lattice vectors $g$ (in $\mbox{\AA}^{-1}$) 
			
			\item {\fontfamily{pcr}\selectfont sg\_max}: defines the maximum excitation error (in $\mbox{\AA}^{-1}$) for which diffracted beams contribute to the Bloch wave calculations. It is also used in the indexing step of multislice.

			\item {\fontfamily{pcr}\selectfont g\_max\_store}: specifies the maximum length of reciprocal lattice vectors for which the diffracted beam intensities are stored.
			
		\end{enumerate}

\subsubsection{Intensity calculation}		
Reflection intensities are computed for all specified orientations using either the Bloch wave or multislice method. For multislice, an additional step of integration of reflection spots is performed. The decision which reflections are integrated is based on the parameters $g_{\mathrm{max}}$ and $S_{g_{\mathrm{max}}}$. Thus, if Bloch wave and multislice calculations are performed, the resulting data set has an identical set of reflections. Likewise, the slice thickness and number of exit planes are used in the Bloch wave calculations to determine for which thicknesses diffracted intensities are calculated. The resulting datasets of diffracted intensities as a function or reflection index, crystal thickness, and orientation are stored in the Zarr format. Note that diffraction patterns calculated on the fly with the multislice method are not stored.

\subsection{Convergence tests} \label{SI_convergence_parameters}

\subsubsection{Tested parameter range}
Here, a list of the parameter values used for the convergence tests is given. Example input files are provided in the github repository.

		\begin{enumerate}
			\item Bloch wave parameter $g_{\mathrm{max}}$ ("{\fontfamily{pcr}\selectfont g\_max}"): 2.0, 2.2, 2.4, 2.6, 2.8, 3.0, 3.5, 4.0, 4.5, 5.0, 5.5,  6.0, 6.5, 7.0, 8.0, 9.0 
			
			\item Bloch wave parameter $S_{g_{\mathrm{max}}}$ ("{\fontfamily{pcr}\selectfont sg\_max}"): 0.01, 0.05, 0.1, 0.2, 0.3, 0.4, 0.5, 0.6, 0.7, 0.8, 0.9, 1.0, 1.5, 2.0, 2.5, 3.0, 4.0

			\item multislice parameter \textbf{box size} ("{\fontfamily{pcr}\selectfont box\_size\_x", "box\_size\_y}"): 25, 33, 40, 50, 54.305, 58, 62.1, 64.3, 70, 75, 80, 90, 92, 100, 103.1795, 10, 114, 114.0405, 120, 125, 130, 145, 148, 150, 160, 162.915, 175, 200, 225, 250

			\item multislice parameter \textbf{slice thickness} ("{\fontfamily{pcr}\selectfont slice\_thickness}"): 0.02, 0.04, 0.05, 0.1, 0.2, 0.25, 0.4, 0.5, 0.8, 1.0, 1.6, 2.0, 4.0

			\item multislice parameter \textbf{sampling} ("{\fontfamily{pcr}\selectfont sampling}"): 0.01, 0.02, 0.03, 0.04, 0.05, 0.06, 0.07, 0.08, 0.09, 0.10, 0.12, 0.14, 0.16, 0.18, 0.2, 0.25, 0.3, 0.4
		\end{enumerate}

\subsubsection{Average scattered intensity} \label{Average_scattered_intensity}

In the article, the parameter $\textrm{RL2}$ related to the root mean square deviation is used to assess the convergence of simulation parameters. There are many alternative quantifiers. Here, we show plots that compare average scattered intensities for a complete dataset, i.e., the average of all reflection intensity values (without $000$) of all orientations $\alpha$ and thicknesses $z$. We call the resulting quantity $\textrm{ASI}_{p}$ (Eq.~\ref{ASI_p}). Likewise, the average of all reflection intensity values (without $000$) of thicknesses $z$ for a given orientation is called $\textrm{ASI}_{p,\alpha}$ (Eq.~\ref{ASI_palpha}). In addition to $\textrm{RL2}$ , the $\textrm{ASI}$ may be further used to select suitable starting parameters for simulations.

\begin{equation} \label{ASI_p}
	\textrm{ASI}_{p} = \langle \sum_{\mathbf{h}} I_{\mathbf{h}}(z,\alpha,p) \rangle = \frac{1}{N_z N_{\alpha} }\sum_z \sum_\alpha \sum_{\mathbf{h}} I_{\mathbf{h}}(z,\alpha,p)
\end{equation}

\begin{equation} \label{ASI_palpha}
\textrm{ASI}_{p,\alpha} = \frac{1}{N_z}\sum_z \sum_{\mathbf{h}} I_{\mathbf{h}}(z,\alpha,p) 
\end{equation}

%\begin{figure}[!h] %
%	\begin{center}
%		\includegraphics[width=0.85\textwidth]{Figures/ASI.png} %%%%%%%%%%%%%%%%%%%%%%%%%%%%%%%%%%%%%%%%%%%%%%%%% enerate PDF/EPS before final publication of SI
%	\end{center}
%	\caption{Average scattered intensity as a function of parameter value normalised by the $\textrm{ASI}$ of the most expensive calculation in the series so that the curves approach a value of 1.} 
%	\label{fig:ASI}
%\end{figure}

\begin{figure}[!h]
    \centering
    \begin{minipage}[b]{0.48\textwidth}
        \centering
        \includegraphics[width=\linewidth]{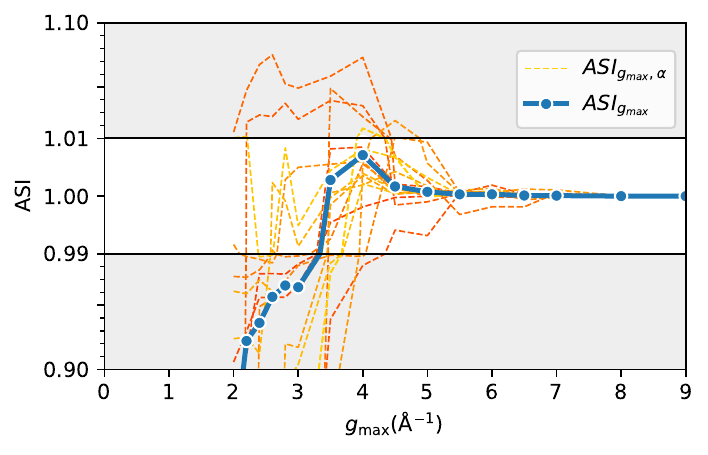}
    \end{minipage}
    \hfill
    \begin{minipage}[b]{0.48\textwidth}
        \centering
        \includegraphics[width=\linewidth]{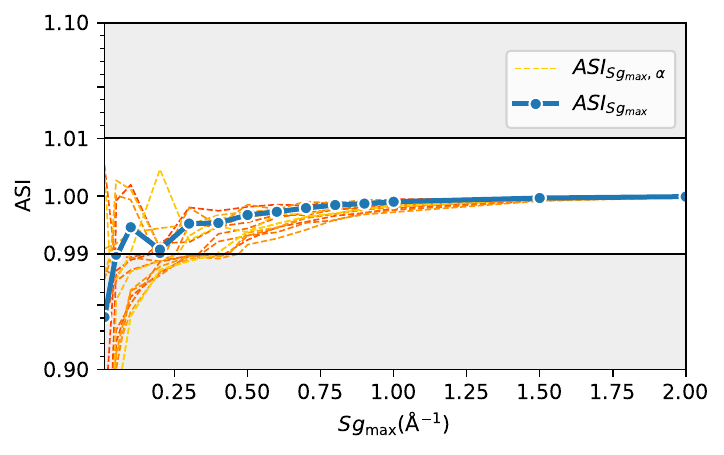}
    \end{minipage}

    \vspace{0.8em}

    \begin{minipage}[b]{0.48\textwidth}
        \centering
        \includegraphics[width=\linewidth]{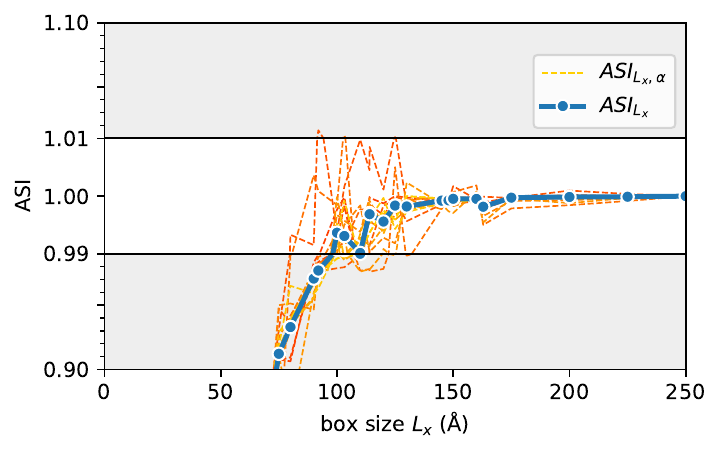}
    \end{minipage}
    \hfill
    \begin{minipage}[b]{0.48\textwidth}
        \centering
        \includegraphics[width=\linewidth]{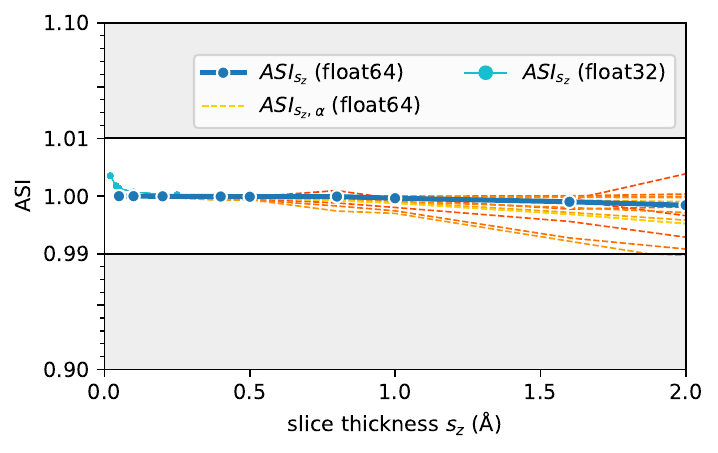}
    \end{minipage}

    \vspace{0.8em}

    \begin{minipage}[b]{0.48\textwidth}
        \centering
        \includegraphics[width=\linewidth]{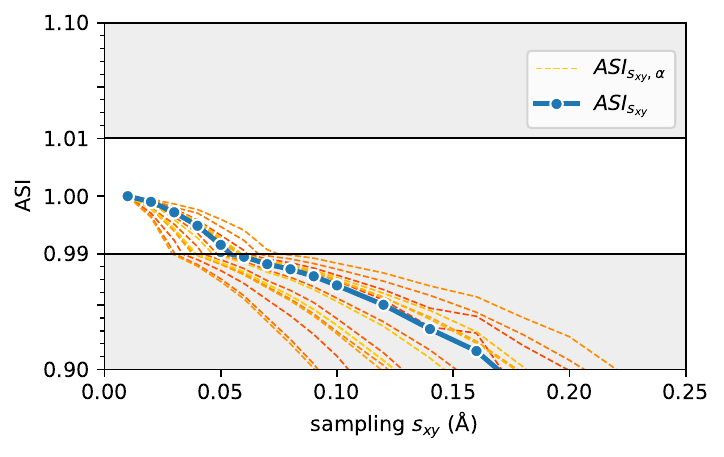}
    \end{minipage}

    \caption{Average scattered intensity as a function of parameter value normalised by the $\textrm{ASI}$ of the most expensive calculation in the series so that the curves approach a value of 1.}
    \label{fig:ASI}
\end{figure}

\subsubsection{Projection method for slice generation} \label{Projection_method}

To generate a 2D slice of electrostatic potential from a distribution of atoms, the typical approach is to first identify all atoms where the atom center is within the slice boundaries and then project the entire electrostatic potential of each of these atoms onto the respective 2D slice. This approach is called "infinite projection" because the electrostatic potential of the atom is integrated over the radius from the center to infinity. In this case, the actual radius of the atom does not matter and numerical noise may decide whether an atom is assigned to one slice or the other. The more accurate approach is to distribute the electrostatic potential of each atom among those slices where the electrostatic potential of that atom is above a certain threshold. This approach is computationally much more expensive because the coordinates are not sufficient to decide to which slices an atom belongs. 

If one generated the entire electrostatic potential of a box of $100~\mbox{\AA} \times 100~\mbox{\AA} \times 1000~\mbox{\AA}$ with a sampling of $0.05~\mbox{\AA}$ and a slice thickness of $0.5~\mbox{\AA}$, with single-precision this would correspond to about 30~GB only to store the electrostatic potential (for one orientation). Even though we expect that there is room for improvement, with the current implementation in \emph{ab}TEM the calculation of 16 orientations of Si took about 16 times longer than the calculations with the infinite projection method. The factor of 16 is based on a range of calculations with various slice thicknesses. Extrapolation suggests that the simulation of one typical 3D ED data set would take several days or several weeks with a GPU-powered desktop computer. The figure below indicates the wall-clock time of the simulation for the convergence check on the slice thickness for 16 orientations of Si as presented in the main text.

\begin{figure}[!h] %
	\begin{center}
		\includegraphics[width=0.75\textwidth]{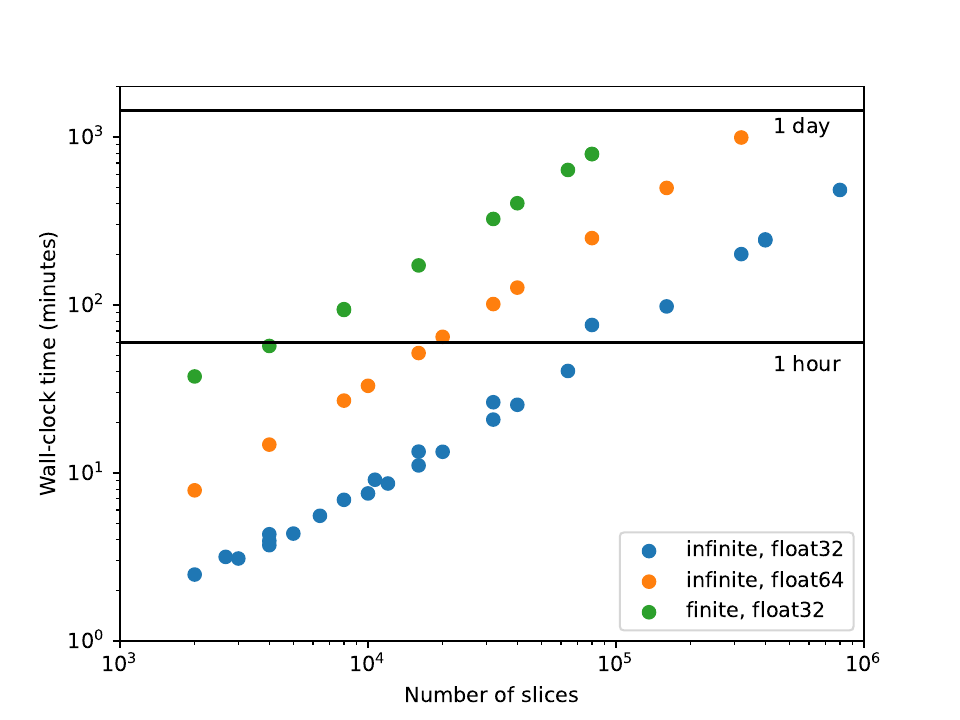}
	\end{center}
	\caption{Comparison of wall-clock time for the simulation of data sets with 16 orientations using multislice. Each point represents one data set, where different slice thicknesses were used. "infinite" indicates that the infinite projection approach was used to calculate the electrostatic potential of 2D slices. "finite" indicates that the finite projection approach was used. A complete data set with 451 orientations, a crystal thickness of 1000~\mbox{\AA} and a slice thickness of 0.5~\mbox{\AA} is based on the computation of about $10^6$ slices.}
	\label{fig:Walltime}
\end{figure}

The $R_{z}$ was calculated for a comparison of corresponding Bloch wave calculations with the multislice data generated with a slice thickness of $0.4~\mbox{\AA}$. For the sake of completeness, data with both single-precision ("float32") and double-precision ("float64") are included. The best agreement for most thickness ranges is achieved for the more accurate finite projection. Double-precision calculations perform better than single-precision calculations, though the difference is negligible. It needs to be emphasised that these results are based on only 16 orientations and that there are no complete rocking curves in these data sets. Therefore, this is considered a preliminary result indicating a trend, which is expected to be qualitatively representative for the simulation of more complete 3D ED data sets. Also note that the agreement factors are visibly better than those presented in the main article because of the coarse angular sampling and are not comparable with $R_{z}$ values presented in the main text, which are all based on 451 orientations involving complete rocking curves and many more reflections.

\begin{figure}[!h] %
	\begin{center}
		\includegraphics[width=0.75\textwidth]{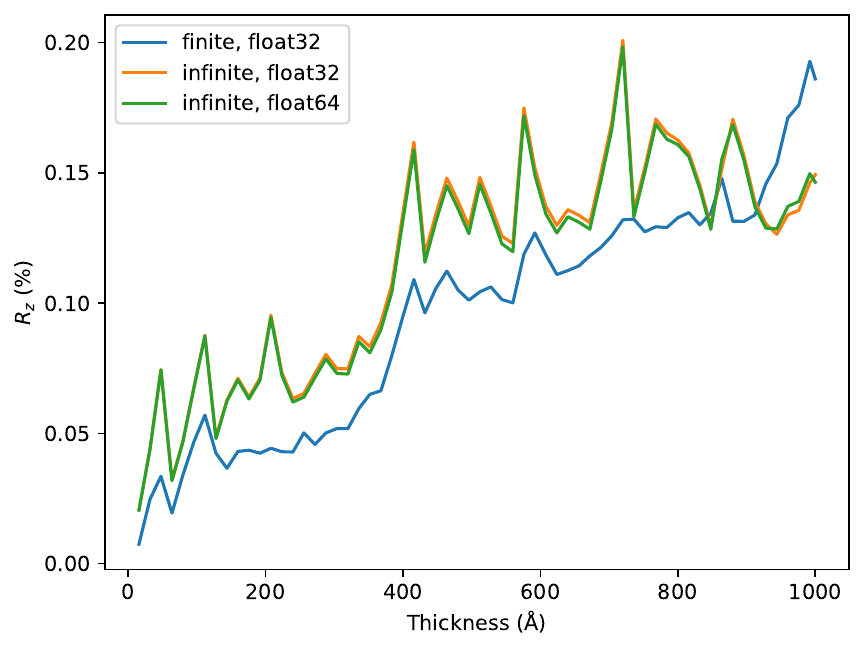}
	\end{center}
	\caption{$R_{z}$ calculated for data sets with 16 orientations.} 
	\label{fig:Sixteen}
\end{figure}

%%% begin Gemini edit

\subsection{Influence of the interaction strength on the agreement between Bloch wave and multislice calculations} \label{SI_interaction_strength}

This section is an extension of Section~\ref{sec:Comparisons} in the main text. The agreement between Bloch wave and multislice calculations was assessed as a function of various parameters, which are known to influence the strength of the electron-matter interaction. Multislice simulations of thicker crystals require more slices, and each propagation step accumulates minor deviations depending on how well simulation parameters converged. Lower kinetic energies increase the interaction strength, and in certain orietations the shortened extinction distances affect intensity oscillations. Heavier atoms increase electrostatic potentials which again lead to a stronger interaction. Together, these factors enhance method discrepancies, underscoring the need for rigorous convergence testing across parameters.

\subsubsection{Dependence on density} \label{Dependence_on_density}

\begin{figure}[!h] %
    \begin{center}
        \includegraphics[width=0.75\textwidth]{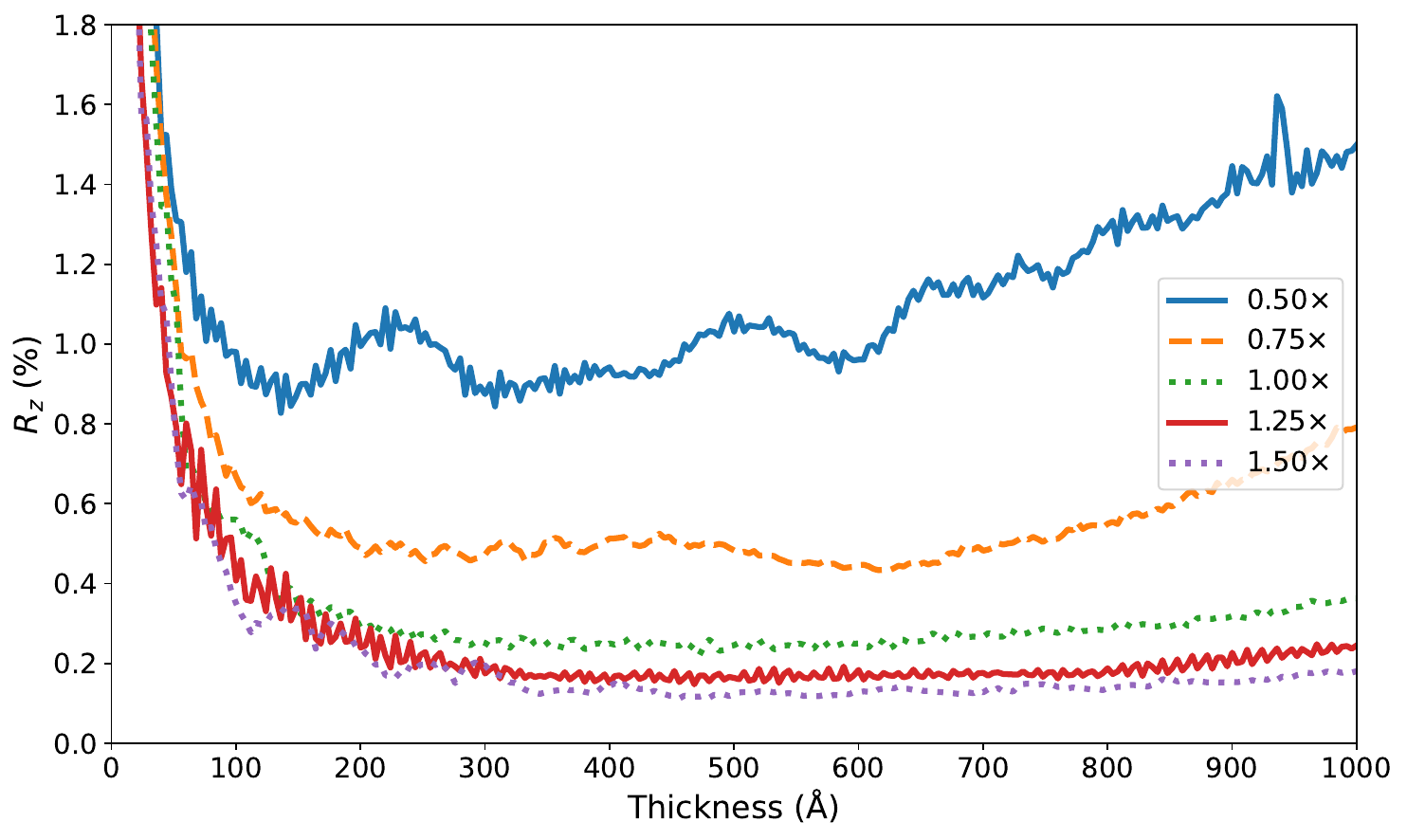}
    \end{center}
    \caption{$R_z$ as a function of crystal thickness for a series of Si structures with unit cell dimensions modified to $0.50\times$, $0.75\times$, $1.00\times$, $1.25\times$, and $1.50\times$ of the original unit cell.} 
	\label{fig:R_vs_cell_size}
\end{figure}

Materials with higher density usually exhibit a larger electrostatic potential, which leads to stronger electron-matter interaction. In this example of artificial structures, we  modified the density without changing the chemical composition. The starting point was a cubic silicon crystal structure. By scaling the lattice parameters to $0.5\times$, $0.75\times$, $1.25\times$, and $1.5\times$ of the original unit cell, we created four configurations with varying densities. As shown in Fig.~\ref{fig:R_vs_cell_size}, $R_z$ increases systematically as the unit cell volume decreases.

\subsubsection{Dependence on kinetic energy and simulation box size} \label{Dependence_on_keV}

\begin{figure}[!h] %
    \begin{center}
        \includegraphics[width=0.95\textwidth]{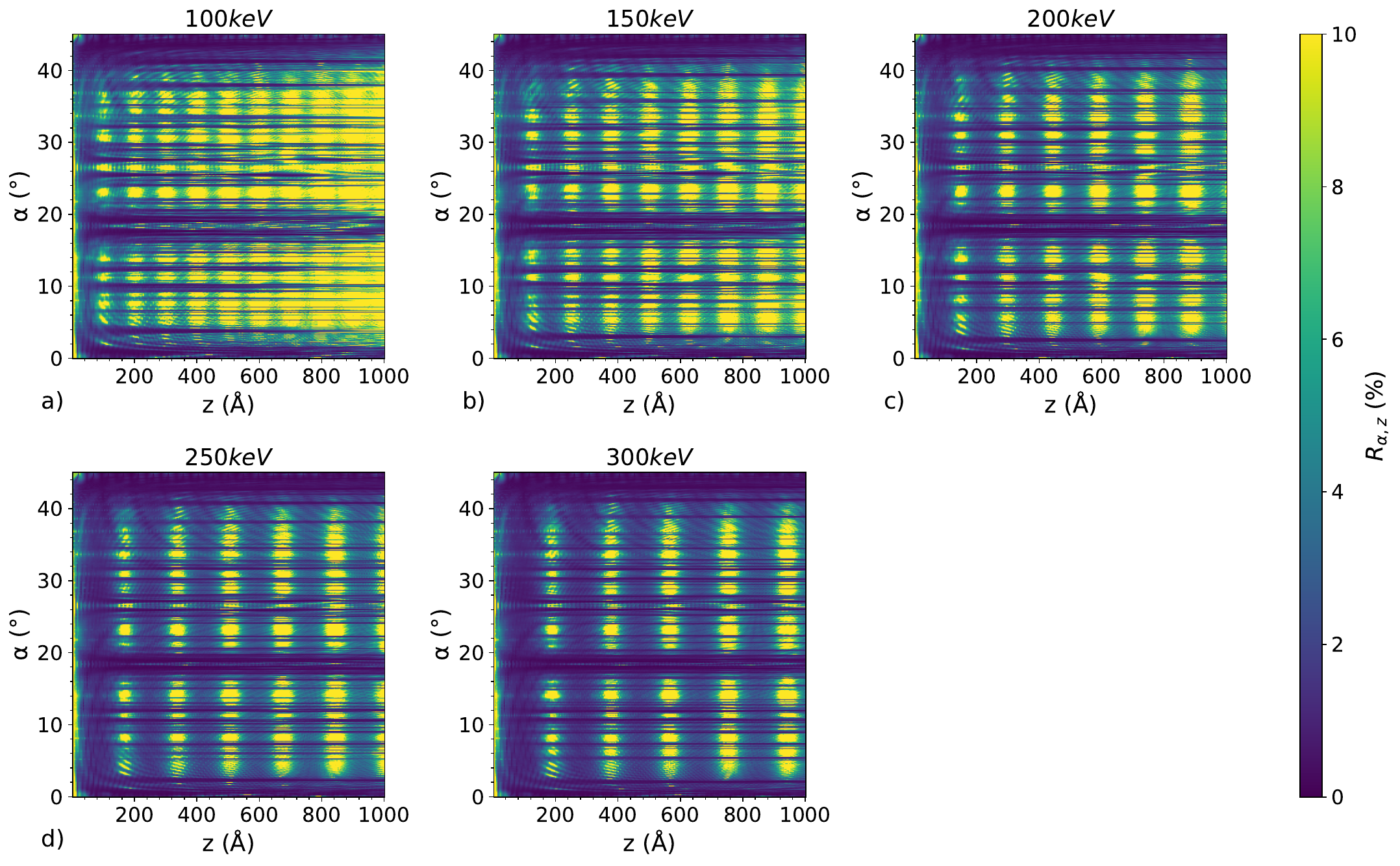}
    \end{center}
    \caption{$R_{z,\alpha}$ as a function of crystal thickness and rotation angle for electron kinetic energies ranging from $100$ to $300~\mbox{keV}$: a) $100$, b) $150$, c) $200$, d) $250$, e) $300~\mbox{keV}$.} %Dark blue represents low $R_z$ and yellow represents high $R_z$.} 
	\label{fig:R_vs_keV}
\end{figure}

Lower kinetic energies result in longer electron wavelengths and larger interaction constants, yielding stronger interaction with the crystal potential. As shown in Fig.~\ref{fig:Energy} in the main text, increasing the kinetic energy from $100$ to $300~\mbox{keV}$ leads to better $R_z$ across all crystal thicknesses. However, the improvement is most pronounced at higher thickness values. Fig.~\ref{fig:R_vs_keV} shows $R_{z,\alpha}$, which indicates that the agreement between multislice and Bloch wave calculations shows no visible dependence on the orientation, but there is a pronounced dependence on the thickness. Due to the special orientation, the strong reflections $400$ and $-400$ are very close to the projection of the rotation axis. Their extinction lengths dominate the overall behavior of the $R_z$ values and agreement factors are poor if almost the entire intensity is concentrated in the 000 reflection. In general, the agreement between the two simulation approaches is worse at larger thickness values, which is more pronounced for lower kinetic energies. 

\begin{figure}[!h] %
    \begin{center}
        \includegraphics[width=0.75\textwidth]{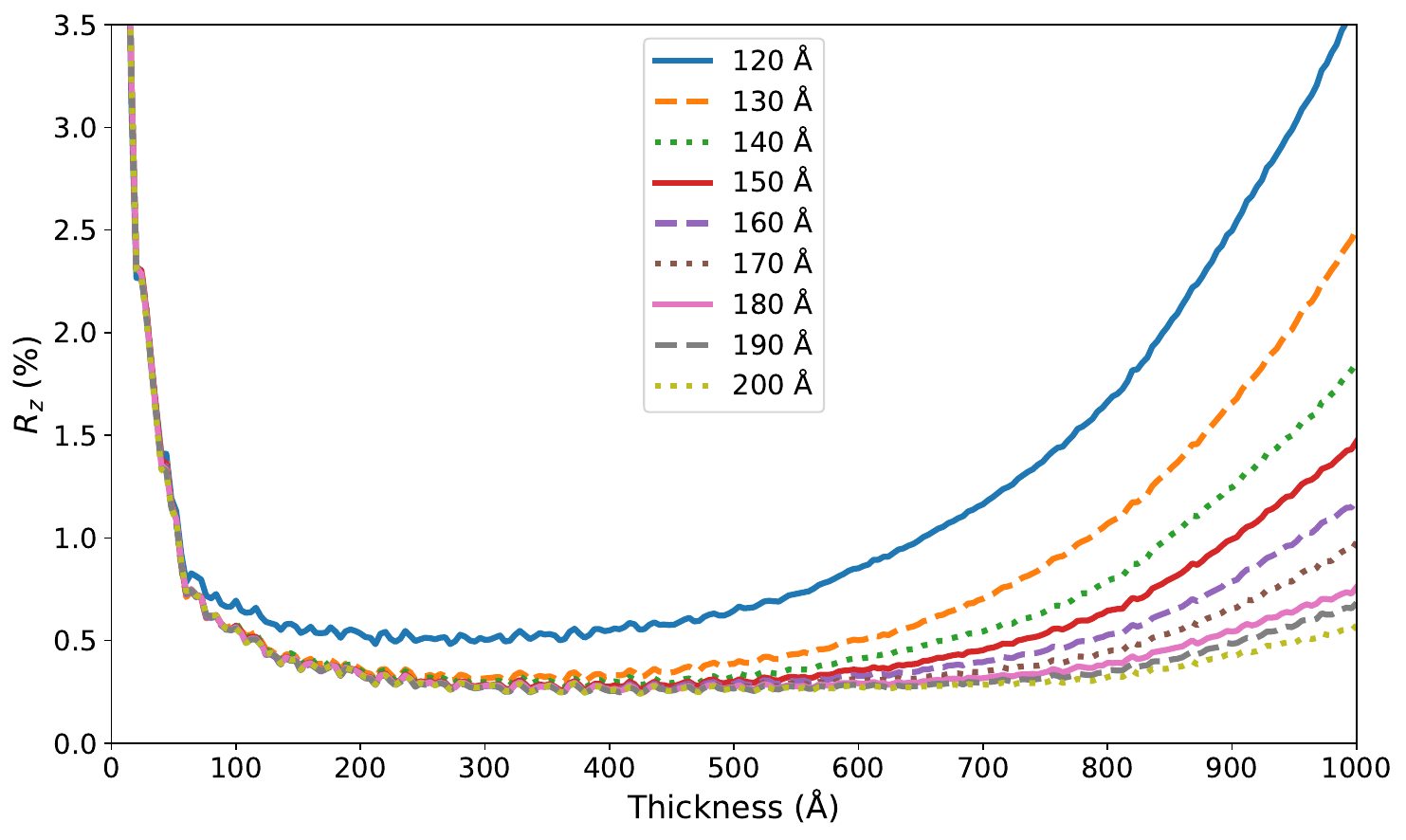}
    \end{center}
    \caption{$R_z$ (\%) as a function of crystal thickness (\mbox{\AA}) at $100~\mbox{keV}$ beam energy for various simulation box dimensions ($L_{x} = L_{y}$).} 
	\label{fig:100keV}
\end{figure}

To test if more stringent convergence parameters can improve the agreement, we performed additional multislice calculations with a larger simulation box sizes for the $100~\mbox{keV}$ case (Fig.~\ref{fig:100keV}). The results show that increasing the box size significantly improves the agreement. Hence, stronger interaction requires more stringent convergence parameters to achieve a better agreement between the two simulation approaches. This also means that the multislice simulations employing lower kinetic energies are computationally more expensive than simulations employing higher kinetic energies.

\subsubsection{Dependence on atomic number} \label{SI:Dependence_on_Z}

\begin{figure}[!h] %
    \begin{center}
        \includegraphics[width=\textwidth]{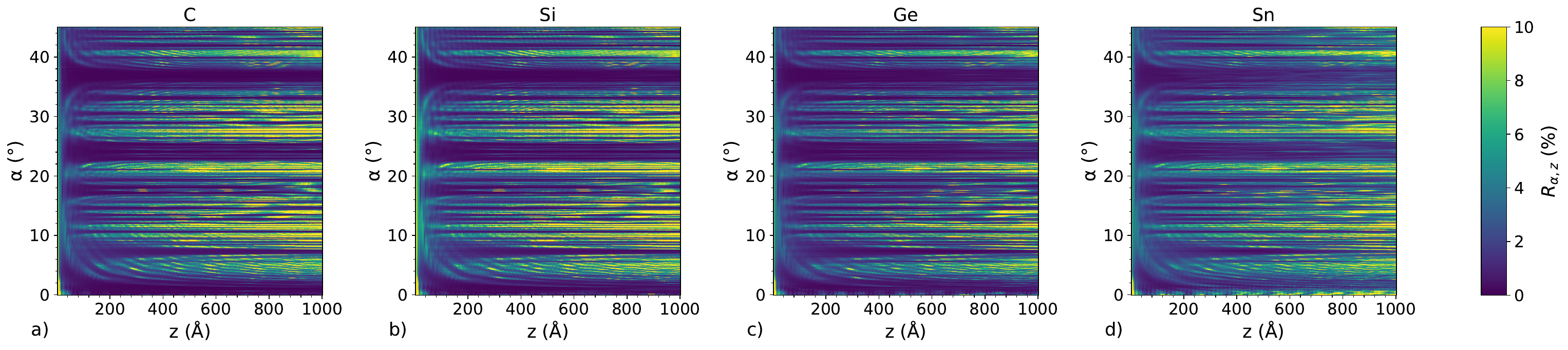}
    \end{center}
    \caption{$R_{z,\alpha}$ as a function of crystal thickness and rotation angle for four diamond-cubic structures substituted with a) carbon (C), b) silicon (Si), c) germanium (Ge), and d) tin (Sn). $\omega$ is set to $25^\circ$ for all structures.} 
	\label{fig:R_vs_chemical_composition}
\end{figure}

As shown in Fig.~\ref{fig:R_vs_chemical_composition}, $R_z$ correlates directly with $Z$. Stronger scattering in high-$Z$ structures increase the sensitivity to parameter convergence, which leads to larger $R_z$ values between the two simulation approaches.

%%%% end Gemini edit

\subsection{Calculation of internal \emph{R}-factors} \label{Calculation_Rint} % $R_{\mathrm{int}}$

To quantify the internal consistency of the simulated data and assess the validity of the kinematic approximation as a function of crystal thickness, the internal residual factor $R_{\mathrm{int}}$ was calculated. Within the kinemtical theory of diffraction, diffracted intensities obey Friedel's law and the point group symmetry of the crystal, resulting in identical intensities reflections that are symetrically equivalent within the Lause class of the crystal structure.
The internal $R$-factor was calculated using the standard crystallographic definition:
$$R_{\mathrm{int}} = \frac{\sum_{\mathbf{h}} \sum_{i} \vert{} I_i(\mathbf{h}) - \langle I(\mathbf{h}) \rangle \vert{}}{\sum_{\mathbf{h}} \sum_{i} I_i(\mathbf{h})}$$

where $\mathbf{h}$ represents a unique reflection index, and $i$ iterates over all symmetrically equivalent observations of that reflection. Intensities $I_i(\mathbf{h})$ are the simulated integrated intensity, and $\langle I(\mathbf{h}) \rangle$ is the mean integrated intensity of a group of symmetically equivalent reflections.

To ensure a robust and physically meaningful calculation of $R_{\mathrm{int}}$, the selection of reflections was filtered based on the following three criteria:

\begin{itemize}
\item Complete rocking curves: Only reflections whose intensity profiles are complete within the simulated rotation range were included. Partial reflections, which are cut off by the start or end of the rotation series, were excluded.

\item Laue class symmetry: Symmetry equivalence was determined according to the ideal Laue class of the respective crystal structure.

\item Exclusion of singletons: The calculation was restricted to groups of symmetrically equivalent reflections that contained at least two fully recorded reflection profiles ($N \geq 2$). Singletons trivially yield $\vert{} I - \langle I \rangle \vert{} = 0$ and were omitted.
\end{itemize}

\subsection{Realistic cases} \label{Realistic_cases}

\begin{figure}[!h] %
	\begin{center}
		\includegraphics[width=\textwidth]{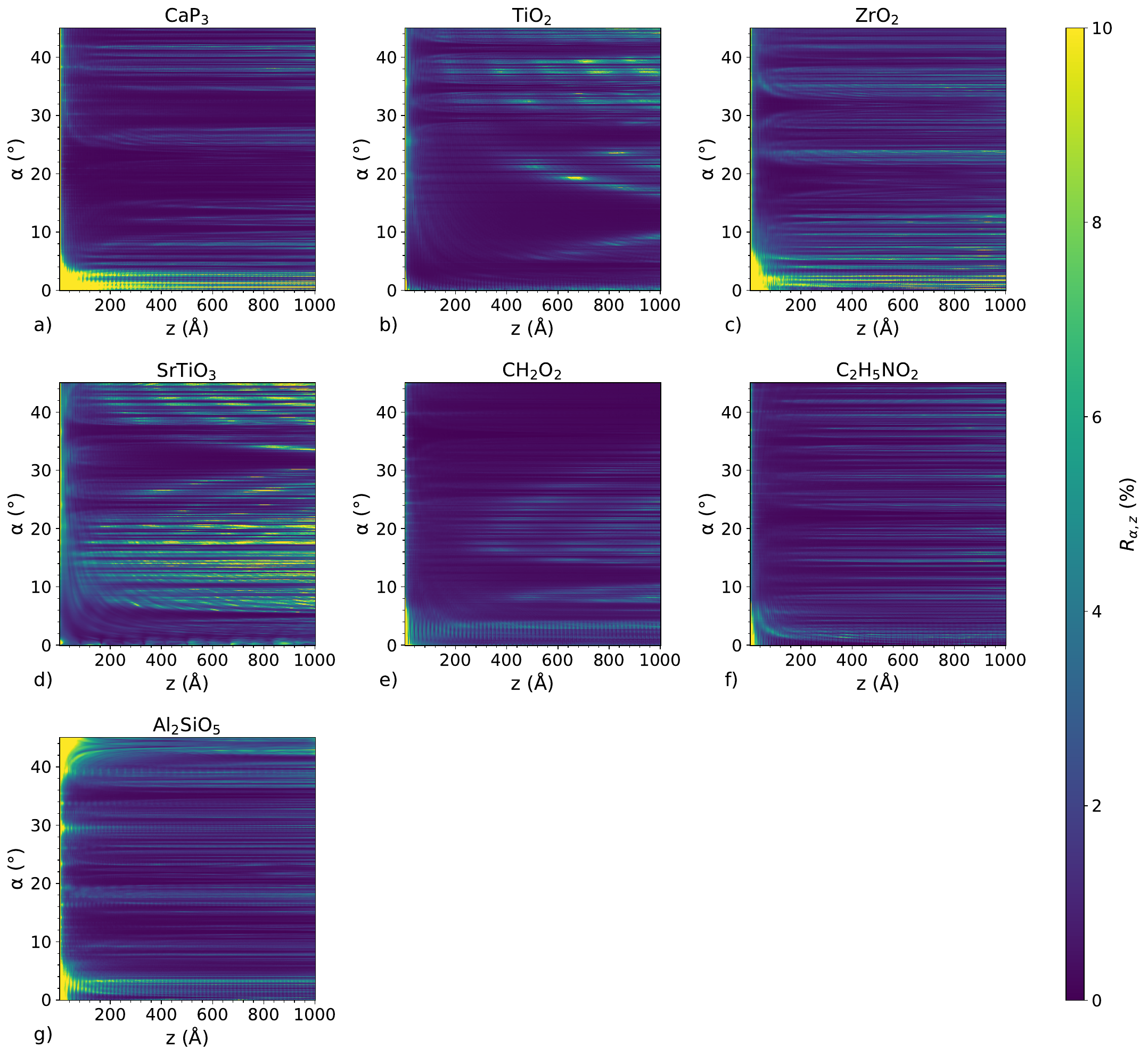} %%%%%%%%
	\end{center}
	\caption{The heatmaps show $R_{z,\alpha}$ as a function of crystal thickness and rotation angle for seven compounds.} 
	\label{fig:Realistic_cases2}
\end{figure}

To show the general applicability and robustness of the pipeline, we apply our comparative analysis to a selection of seven real crystal structures, each presenting some challenges for electron diffraction simulations. As shown in Fig.~\ref{fig:Realistic_cases2}, the agreement between the Bloch wave and multislice methods remains high across most systems, especially for the organic crystals (formic acid and glycine) where the low scattering power of the light atoms.

\subsection{Windowing} \label{Windowing}

The application of a 2D window on the exit wave introduces a systematic modification of the amplitudes of the Fourier-transformed image. In the case of a 2D Hann window, the mean value of the window function $\left< g(x,y) \right>$ is $(1/2)^2$. Therefore, Fourier amplitudes are reduced by the same factor of $1/4$. The intensities of the diffraction pattern, i.e., of the squared amplitudes, are modified by the mean of the squared window function $\left< g(x,y)^2 \right> = (3/8)^2$. Therefore, diffracted intensities calculated with multislice using a Hann window as described need to be multiplied by a factor $(8/3)^2 \approx 7.11$.

Windowing has a striking effect on the agreement between multislice and Bloch wave results. For the triclinic variant of Si with unit cell parameters of $5.3$, $5.8$, and $5.4~\mbox{\AA}$, and interaxial angles of $80$, $95.4$, and $101^\circ$, another data set of multislice was generated, this time without applying a windowing function. The mean $\left< R_{z} \right>$ increased from $1.1 \%$ ("windowing": "hann") to $5.0 \%$ (without windowing). 

\subsection{Technical aspects}
The computational requirements, efficiency and runtime of 3D ED simulations are dependent on several technical factors, including the numerical precision, data structures, and hardware architecture. Here we wish to discuss a few selected aspects.

In general, double-precision floating-point numbers (64~bits) are used to represent real numbers. Imaginary numbers are based on two real numbers, each with double precision totalling to 128~bits. All quantities related to Bloch wave calculations use this numerical precision, even if data are stored in single precision. Multislice calculations by default use single-precision floating-point numbers (32~bits per real number) for the electrostatic potential, the wave function, and diffraction patterns. This reduction in precision for multislice is a practice meant to optimize memory and computation power usage, as the iterative nature of the algorithm is typically less sensitive to the precision loss than the matrix diagonalization required in Bloch wave calculations.

To handle the large volumes of data generated by these simulations (5–10~GB and more), particularly when spanning multiple orientations and thicknesses, we utilize the Zarr data format. Zarr provides a chunked, compressed, and N-dimensional array storage that is highly efficient for parallel I/O operations. Intensity values in output files can be written either as single-precision or double-precision floating-point numbers. By storing results in Zarr format, we ensure that a relevant subset of the data can be accessed and analyzed without loading the entire dataset into memory.

The simulations in this work were performed on two devices. First, a high-performance computing cluster of e-INFRA CZ with both Central Processing Units (CPUs) and Graphics Processing Units (GPUs) was used. Second, a desktop computer running on Windows 11 with an AMD Ryzen 9 5950X CPU and an NVIDIA GeForce RTX 3060 GPU was employed. The desktop computer was used for most of the convergence tests, while the high-performance computing cluster was mainly used for the simulations of the realistic cases. We observe a substantial speedup when using GPUs for multislice and Bloch wavecalculations.

Our implementation uses the CUDA-powered capabilities of the \emph{ab}TEM package. For researchers looking to implement this pipeline, we recommend the use of modern GPUs with at least 12~GB of VRAM to handle the large supercells and high-resolution sampling required for converged results. The use of GPU acceleration typically results in a speedup of approximately one order of magnitude compared to CPU-only execution. The results of the convergence tests may serve as a guideline for the choice of simulation parameters. Results with lower accuracy may still be useful for certain applications, such as training data for machine learning, where the computational cost is a critical factor.

With a desktop computer, the simulation of data sets with the Bloch wave approach typically takes several minutes for the smallest unit cells, and around one hour for larger unit cells presented in this article. In contrast, the multislice approach is computationally more expensive, typically requiring approximately one day of runtime.

\begin{acknowledgements}
Computational resources were provided by the e-INFRA CZ project (ID:90254), supported by the Ministry of Education, Youth and Sports of the Czech Republic.
\end{acknowledgements}

\begin{funding}
List funding organizations, recipients, grant numbers, etc.
\end{funding}

\ConflictsOfInterest{There are no conflicts of interest to declare.
}

\DataAvailability{All scripts and a selection of input files used for the calculations presented in this study are available on github (https://github.com/3DED/py3DED/)
}

\bibliography{References} % basename of .bib file

\end{document}